\documentclass[fleqn,usenatbib]{mnras}
\usepackage{newtxtext,newtxmath}

\usepackage[T1]{fontenc}
\usepackage{ae,aecompl}
\usepackage{tabularx}
\usepackage{subfig}
\usepackage{color, soul}

\usepackage{graphicx}	
\usepackage{amsmath}	
\usepackage{amssymb}	
\usepackage{cleveref}

\newcommand{\HI}{H\,{\sevensize I}}

\newcommand{\SiII}{Si{\scriptsize~II}}
\newcommand{\MgII}{Mg{\scriptsize~II}}

\newcommand{\NHI}{$N\rm _{H\,{\sevensize I}}$}

\newcommand{\lya}{Ly$\alpha$}

\newcommand{\msun}{M$_\odot$}

\newcommand{\kms}{$\rm km~s^{-1}$}

\newcommand{\eagle}{{\sc eagle}}
\newcommand{\GALICS}{{\sc galics}}
\crefformat{section}{\S#2#1#3} 
\crefformat{subsection}{\S#2#1#3}
\crefformat{subsubsection}{\S#2#1#3}

\title[MUSE surveys the environments of six DLAs]{Linking gas and galaxies at high redshift: MUSE surveys the environments of six damped Ly$\alpha$ galaxies at $z \approx 3$}

\author[R. Mackenzie et al.]{Ruari Mackenzie,$^{1,2}$\thanks{E-mail: mruari@phys.ethz.ch}
Michele Fumagalli,$^{2,3}$, Tom Theuns,$^{3}$ David J. Hatton,$^{2}$ 
\and  Thibault Garel,$^{4,5}$,   Sebastiano Cantalupo$^{1}$, Lise Christensen$^{6}$, Johan P.~U. Fynbo$^{6}$, 
  \and Nissim Kanekar$^{7}$, Palle M{\o}ller$^{8}$,  John O'Meara$^{9,10}$, J. Xavier Prochaska$^{11,12}$,  
  \and Marc Rafelski$^{13}$, Tom Shanks$^{2}$ , James Trayford$^{14}$
\\
 $^{1}$Department of Physics, ETH Zurich, Wolfgang-Pauli-Strasse 27, 8093 Zurich, Switzerland\\
$^{2}$Centre for Extragalactic Astronomy, Durham University, South Road, Durham, DH1 3LE, UK \\
$^{3}$Institute for Computational Cosmology, Durham University, South Road, Durham, DH1 3LE, UK \\
$^{4}$Univ Lyon, Univ Lyon1, Ens de Lyon, CNRS, Centre de Recherche Astrophysique de Lyon UMR5574, F-69230, Saint-Genis-Laval, France \\
$^{5}$Observatoire de Gen\`eve, Universit\'e de Gen\`eve, 51 Ch. des Maillettes, 1290 Versoix, Switzerland \\
  $^{6}$Dark Cosmology Centre, Niels Bohr Institute, University of Copenhagen, Juliane Maries Vej 30, DK-2100 Copenhagen, Denmark\\
  $^{7}$National Centre for Radio Astrophysics, TIFR, Post Bag 3, Ganeshkhind, Pune 411007, India\\
  $^{8}$European Southern Observatory, Karl-Schwarzschild-Strasse 2, 85748, Garching, Germany\\
  $^{9}$Department of Physics, St. Michael's College, Colchester, VT 05439, USA\\
  $^{10}$W. M. Keck Observatory, Waimea, HI 96743, USA\\
  $^{11}$Department of Astronomy and Astrophysics, University of California, 1156 High Street, 
  Santa Cruz, CA 95064 USA \\
  $^{12}$University of California Observatories, Lick Observatory 1156 High Street, Santa Cruz, 
  CA 95064 USA \\
  $^{13}$Space Telescope Science Institute, Baltimore, MD, USA \\
  $^{14}$Leiden Observatory, Leiden University, PO Box 9513, NL-2300 RA Leiden, the Netherlands \\
}

\pubyear{\the\year}

\hypersetup{draft}
\begin{document}
\label{firstpage}
\pagerange{\pageref{firstpage}--\pageref{lastpage}}
\maketitle

\begin{abstract}
We present results from a survey of galaxies in the fields of six $z\ge 3$ Damped Lyman $\alpha$ systems (DLAs) using the Multi Unit Spectroscopic Explorer (MUSE) at the Very Large Telescope (VLT). We report a high detection rate of up to $\approx 80\%$ of galaxies within 1000 km/s from DLAs and with impact parameters between 25 and 280 kpc. In particular, we discovered  5 high-confidence Ly$\alpha$ emitters associated with three DLAs, plus up to 9 additional detections across five of the six fields. The majority of the detections are at relatively large impact parameters ($>50$ kpc) with two detections being plausible host galaxies. Among our detections, we report four galaxies associated with the most metal-poor DLA in our sample ($Z/Z_\odot = -2.33\pm0.22$), which trace an overdense structure resembling a filament. By comparing our detections with predictions from the Evolution and Assembly of GaLaxies and their Environments (EAGLE) cosmological simulations and a semi-analytic model designed to reproduce the observed bias of DLAs at $z>2$, we conclude that our observations are consistent with a scenario in which a significant fraction of DLAs trace the neutral regions within halos with a characteristic mass of $M_{\rm h} \approx 10^{11}-10^{12}~\rm M_\odot$, in agreement with the inference made from the large-scale clustering of DLAs. 
We finally show how larger surveys targeting $\approx 25$ absorbers have the potential of constraining the characteristic masses of halos hosting high-redshift DLAs with sufficient accuracy to discriminate between different models.   
\end{abstract}




\section{Introduction}

Since the discovery of damped \lya\ absorbers (DLAs) in the 
spectra of quasars in the 1970s \citep{beaver1972,carswell1975}, significant 
efforts have been made to identify the properties of 
the galaxy population that gives rise to this class 
of absorption line systems. The interest in connecting
DLAs to galaxies stems from the fact that these absorbers, 
defined to have neutral hydrogen column density in excess 
of $\log (N_{\rm HI}/\rm cm^{-2}) \ge 20.3$ \citep{wolfe2005}, act 
as signposts of significant reservoirs of neutral hydrogen 
within or around high-redshift galaxies \citep{wolfe1986}.
For this reason, direct associations between DLAs detected in absorption and galaxies detected in emission provide a powerful way to probe links between the gas supply, in the form of neutral hydrogen, and ongoing star formation. The combination of absorption and emission techniques thus provides at present the only means to study the star formation law in atomic gas beyond $z>2$  \citep{wolfe2006,prochaska2009,rafelski2011,fumagalli2015,rafelski2016}. 
 
 Starting from earlier searches from the ground and with the {\it Hubble Space Telescope} \citep{WarrenHST,Moller04}, several surveys have attempted to identify the galaxy population that gives rise to DLAs. Despite decades of searches, progress has been scarce until recently. Building on the evidence that galaxies obey a defined mass-metallicity relation \citep{Tremonti04, Maiolino08}, searches have focused on the high end of the metallicity distribution of DLAs, yielding higher detection rates with the discovery of tens of DLA hosts \citep{Fynbo10,Fynbo2013,Krogager17}. The advent of integral field spectrographs (e.g. SINFONI at the Very Large Telescope, VLT; \citealp{Peroux12} and OSIRIS at the W. M. Keck Observatory; \citealp{Jorgenson2014}) have enabled more efficient spectroscopic follow-up as all of the relevant solid angle around the quasars can be covered in a single setting. The much increased sensitivity of the Atacama Large Millimetre Array (ALMA) is now also enabling searches of DLA galaxies via molecular and atomic lines, a technique that is being successfully pioneered, still, at the high end of the metallicity distribution \citep{neeleman2017,Kanekar18, Fynbo2018, Neeleman2019}. 

While these recent identifications offer a way to finally study the link between column density and metallicity in absorption, and stellar masses and star formation rates (SFRs) in emission \citep{Moller04,Christensen2014,Krogager17}, these studies are likely to only probe the bright end of the DLA population, and may not be fully representative of the diverse population of DLA host galaxies.  
Most simulations and models \citep{Haehnelt1998,Fynbo99, Pontzen2008, Barnes2009,RahmatiSchaye,Bird2014, Fynbo2008} consistently indicate that DLA hosts are generally to be found at the very faint end of the luminosity function, typically below the sensitivity limit of current searches. Indeed, our own survey of galaxies designed to image the DLA hosts at all impact parameters \citep{OMearaSHOT,fumagalli2010,fumagalli2014,fumagalli2015} has yielded a series of non-detections within a few kiloparsecs of the location of the DLAs, despite removing the primary source of observational bias, i.e. the bright quasar emission which hampers the detection of faint galaxies at low impact parameters.

While there seems to be general consensus that DLAs are primarily associated with faint galaxies \cite[e.g.][]{Fynbo99, Krogager17}, the question of what the typical range of halo masses giving rise to DLAs 
remains open. At face value, following a similar argument than the one adopted in abundance matching studies \citep[e.g.][]{ConroySHAM}, DLAs are expected to arise primarily in faint galaxies \citep[e.g.][]{Fynbo2008} and hence in low mass halos ($M_{\rm halo} \lesssim 10^{11} M_\odot$). However, this appears to be in tension with other pieces of evidence. Firstly, the distribution of velocity widths measured from metals in low ionisation states shows a prominent tail at high velocity, which has been used to argue for the existence of a population of large disks hosting DLAs (\citealt{Prochaska1998}; but see \citealt{Bird2015} for more recent work on this topic). Furthermore, \cite{FontRibera12}, using the SDSS-III Baryon Oscillation Spectroscopic Survey (BOSS), found an unexpected large linear bias of DLAs ($b_{DLA} = 2.17\pm0.20$, with an assumption on the bias of the \lya\ forest) by cross-correlating these absorbers with the \lya\ forest. This measurement was more recently updated by \cite{BOSS2018}, who found a linear bias of $b_{DLA} = 2.00 \pm 0.19$, which is only slightly lower than the clustering amplitude of Lyman Break Galaxies \citep[LBGs; see also][]{Cooke2006}. This value of the bias, which implies masses of $\gtrsim 10^{11}~\rm M_\odot$, appears uncomfortably high for some galaxy formation simulations and models (see e.g \citealt{Pontzen2008,Barnes2014, Padmanabhan2017}) if DLAs sample a very wide range in galaxy sizes (e.g. \citealt{Krogager17}). Recently, \cite{PerezRafols18b} have shown that the bias of DLAs has a dependence on metallicity in line with observations and modelling of DLAs which associate more metal-rich DLAs with more massive galaxies \citep{Neeleman2013,Christensen2014}.

Building on our previous searches for galaxies at small impact parameters from the quasars \citep{fumagalli2015}, this study exploits the power of wide field integral field spectroscopy provided by the Multi Unit Spectroscopic Explorer (MUSE) at the VLT \mbox{\citep{BaconMUSE}} to carry out the first highly-complete spectroscopic survey in an sample of DLAs, not selected in metallicity or column density, with the goal of 
searching for faint Ly$\alpha$ emission from galaxies associated with the DLAs. Our survey improves upon previous searches of this type, which were characterised by a lower sensitivity (e.g. \citealp{Christensen07}) or smaller field of view (e.g. \citealp{Peroux12}). By achieving a flux-limited search up to $\approx 250~\rm kpc$ from the DLAs, we are finally able to search at the same time for the DLA hosts and for other associations at larger impact parameters, which provides a means of characterising the DLA environment via the small-scale clustering of galaxies and DLAs. 

Throughout this work, we define a host galaxy as any detection that is physically connected with the absorbing gas. For detected galaxies at small impact parameters, i.e. within a projected distance $b_{\rm impact} \le 50~\rm kpc$, we will consider if they could plausibly be linked to the absorbing material. These values, albeit arbitrary, define a reasonable range of distances and velocities that encompass the inner circumgalactic medium (CGM) of a galaxy. We define more generally an association as
a galaxy that is physically connected to the DLA (e.g. in the same halo or clustered on small scales of up to a few hundred kiloparsecs), but not necessarily the galaxy from which the absorption arises. When comparing our observations with simulations we use a definition of $\Delta v_{\rm DLA, gal} \le 1000 ~\rm km~s^{-1}$ as the definition of associated, a condition that is imposed by the finite volume of the adopted simulations.

This paper is structured in the following way. The observations, data reduction and analysis are presented in Sect. \ref{sec:data_reduct} and \ref{sec:GalSearch}, followed in Sect. \ref{sec:models} by technical details on the simulations and semi-analytic models used in the interpretation. We briefly describe some highlighted detections in Sect. \ref{sec:results}. The discussion of results is presented in Sect. \ref{sec:discussion}, followed by a summary and conclusion in Sect. \ref{sec:conclusion}. Further  discussion of the properties of each field is continued from Sect. \ref{sec:results} in appendix \ref{ap:field_discuss} for the remaining fields. Readers primarily interested in the main results of the paper can focus their attention on Sect. \ref{sec:results}, \ref{sec:discussion} and Sect. \ref{sec:conclusion}.

Magnitudes reported in this paper follow the AB system and fluxes and magnitudes have been corrected for Galactic extinction following \cite{SandFdust}. All spectral data products are reported in vacuum wavelengths, and we adopt the \cite{Planck2015} $\Lambda$CDM cosmological parameters ($\Omega_{\rm m} = 0.308$, $h = 0.677$).

\begin{figure}
	\centering
	\includegraphics[width=0.5\textwidth]{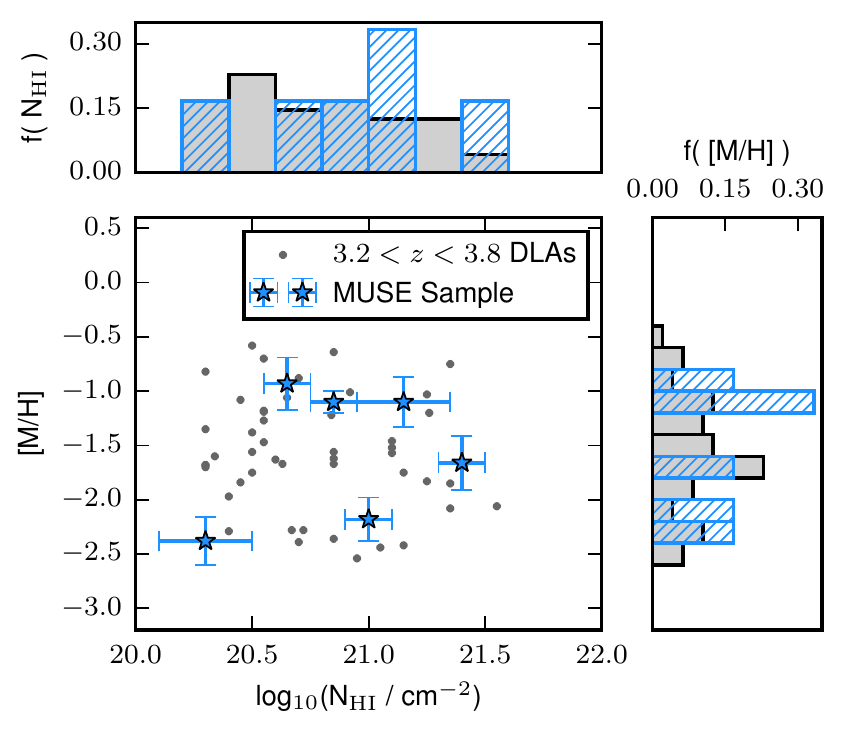}
	\caption{The metallicity and column density distributions of the DLAs targeted with MUSE, compared to the parent DLA population over the same redshift interval ($3.2<z<3.8$) from \protect\cite{rafelski2012}. The normalized distribution of metallicity is shown on the right for both samples, and similarly at the top for column density. In both histograms the MUSE sample are showed in blue hashed, with the parent population in grey behind. Our  sample spans a wide range of DLA properties, thus extending the parameter space covered by most recent surveys.}
	\label{fig:sample_context}
	\vspace{-1em}
\end{figure}

\begin{figure*}
	\centering
	\includegraphics[width=1.0\textwidth]{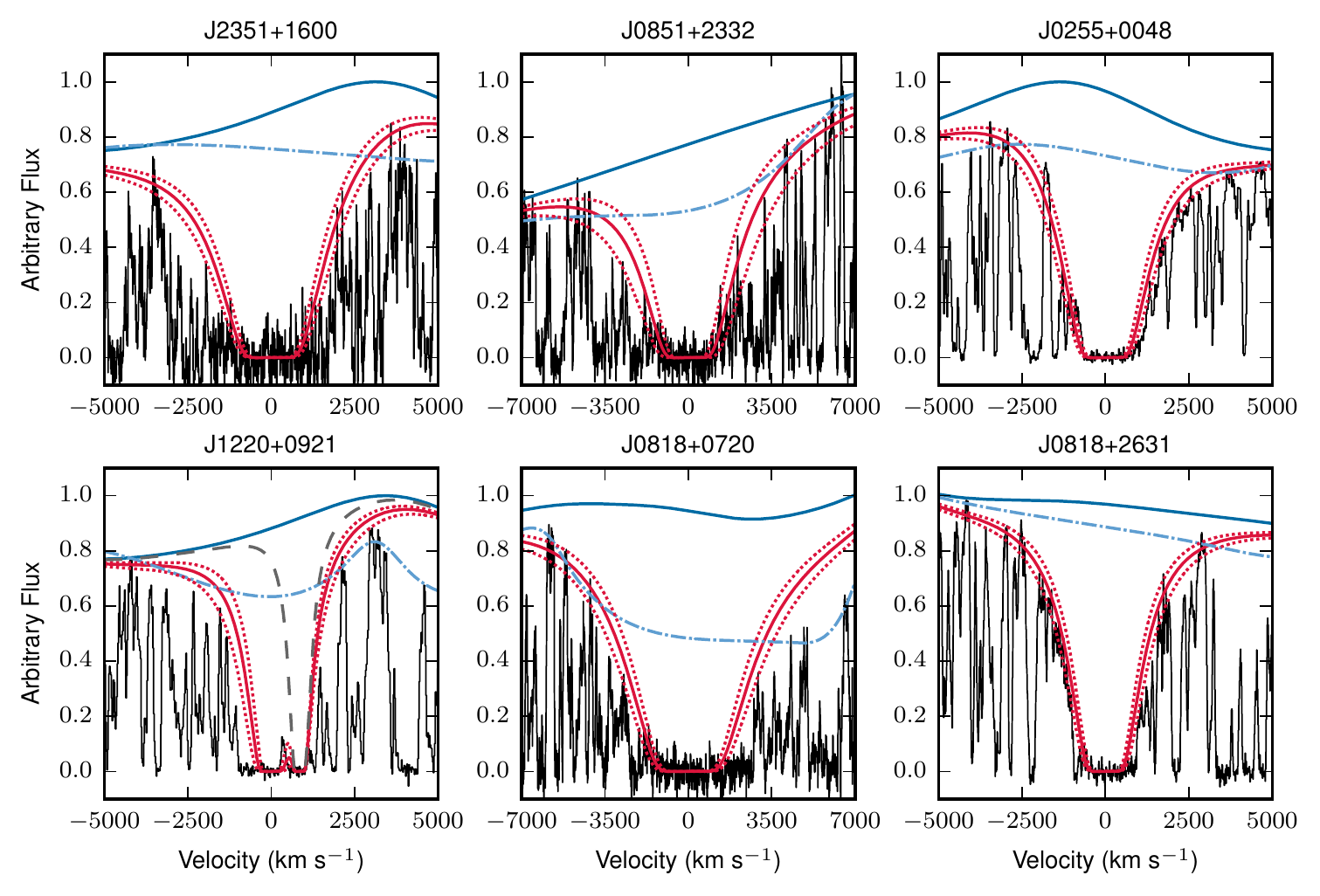}
	\caption{The \lya\ absorption profiles for the six DLAs in the MUSE sample. The quasar spectra are shown in black, the fitted quasar composite template are in blue (solid line), and the template multiplied by the Voigt profile is in red (solid line).  Errors (1$\sigma$) on the column density are show by red dotted lines. The original continuum normalisation from \protect\cite{fumagalli2014} is also shown as a blue dash-dotted line. For J0851+2332 a manual spline continuum normalisation was adopted instead of the quasar composite. The fitted contribution of the Ly$\beta$ line from a proximate DLA contaminating the profile for J1220+0921 is also plotted with grey dashed line.}
	\label{fig:column_densities}
	\vspace{-1em}
\end{figure*}

\section{Observations and data reduction} \label{sec:data_reduct}
\subsection{Sample selection}

The six quasar sightlines observed with MUSE are chosen from the parent sample of \cite{fumagalli2014} which is
selected among the general DLA population purely based on the presence of two optically-thick absorbers
along the quasar sightline, with the goal of searching for the rest-frame UV emission at close impact parameters 
from the DLAs \citep[see][for details]{OMearaSHOT}. 
For this study, we select DLAs that are observable from VLT and at $z >  3.2 $, which is the redshifts at which Ly$\alpha$ falls at wavelengths where the throughput of MUSE is $> 25\%$. Excellent ancillary data, including deep UV and optical imaging and high-resolution spectroscopy are available for this sample, which spans a wide range of metallicity and column density.  

The properties of the selected DLAs in our sample are summarised in Table \ref{tab:sample}, while Figure \ref{fig:sample_context} shows the MUSE sample in context with the wider DLA population in terms of metallicity and column density, highlighting how our targets span a representative range of both parameters. This is in contrast to many recent searches for DLA hosts, such as those conducted with X-Shooter \citep{Krogager17} and ALMA \citep{neeleman2017,Neeleman2019}, which typically select high metallicity systems ($\rm [Si/H] \gtrsim -1$). This selection exploits the metallicity-luminosity relation between the metallicity of the DLA in absorption and the luminosity of the host galaxy to ensure higher detection ratio compared to samples selected, e.g., only on \NHI\ (e.g. \citealp{Peroux12}) or \MgII\ (e.g. \citealp{Bouche2012}). This approach, however, leaves the hosts of the majority of ``typical'' DLAs at $z>2$, with metallicities $\leq5\%$ solar, unexplored.  Our DLA sample contains instead three systems above average metallicity at this redshift \citep{rafelski2012} and three below, thus extending the parameter space targeted by previous searches. Notably, DLA J1220+0921 in our sample is very metal-poor, sitting at $Z/Z_{\odot} = -2.33$ although still somewhat above the metallicity floor at $Z/Z_{\odot} \simeq -3$ \citep{Prochaska2003,rafelski2012}. One sightline in our sample was previously presented in \citealt{fumagalli2017}, the DLA was revealed to be part of a $\simeq50$ kpc structure which may be evidence of an ongoing merger.

\begin{table*}
	\centering
	\begin{tabular}{llccccccc}
		\hline
		QSO Name & Field$^{a}$ & R.A. & Dec & $z_{\rm DLA}$ & log(\NHI / cm$^{-2}$) & log (Z/Z$_\odot$) & Element$^{b}$ & Exp. Time (s)$^{c}$ \\ \hline
		J2351+1600 & 06G6  & 23:51:52.80 & +16:00:48.9 & 3.7861 & 21.00 $\pm$ 0.10 & -2.18 $\pm$ 0.20 & Fe & 6x1480 \\
		J0851+2332 & 10G11 & 08:51:43.72 & +23:32:08.9 & 3.5297 & 21.15 $\pm$ 0.20 & -1.10 $\pm$ 0.23 & Zn  & 4x1480\\
		J0255+0048 & 15H3  & 02:55:18.58 & +00:48:47.6 & 3.2530 & 20.85 $\pm$ 0.10 & -1.10 $\pm$ 0.10 & Si  & 6x1480 \\
		J1220+0921 & 19H7  & 12:20:21.39 & +09:21:35.7 & 3.3090 & 20.30 $\pm$ 0.20 & -2.33 $\pm$ 0.22 & Si  & 6x1480 \\
		J0818+0720 & 23H11 & 08:18:13.14 & +07:20:54.9 & 3.2332 & 21.15 $\pm$ 0.10 & -1.66 $\pm$ 0.25 & Si,Zn  & 6x1480 \\
		J0818+2631 & 24H12 & 08:18:13.05 & +26:31:36.9 & 3.5629 & 20.65 $\pm$ 0.10 & -0.93 $\pm$ 0.24 & Si,Zn  & 2x1480 \\ \hline
	\end{tabular}
	\caption{The properties of the MUSE DLA sample.$^{a}$ The naming system from \citealp{fumagalli2014}. $^{b}$ The element(s) used to estimate the DLA metallicity. $^{c}$ The exposure time of the MUSE integral field spectroscopy. }
	\label{tab:sample}
\end{table*}

\begin{figure*}
	\centering
	\includegraphics[width=1.0\textwidth]{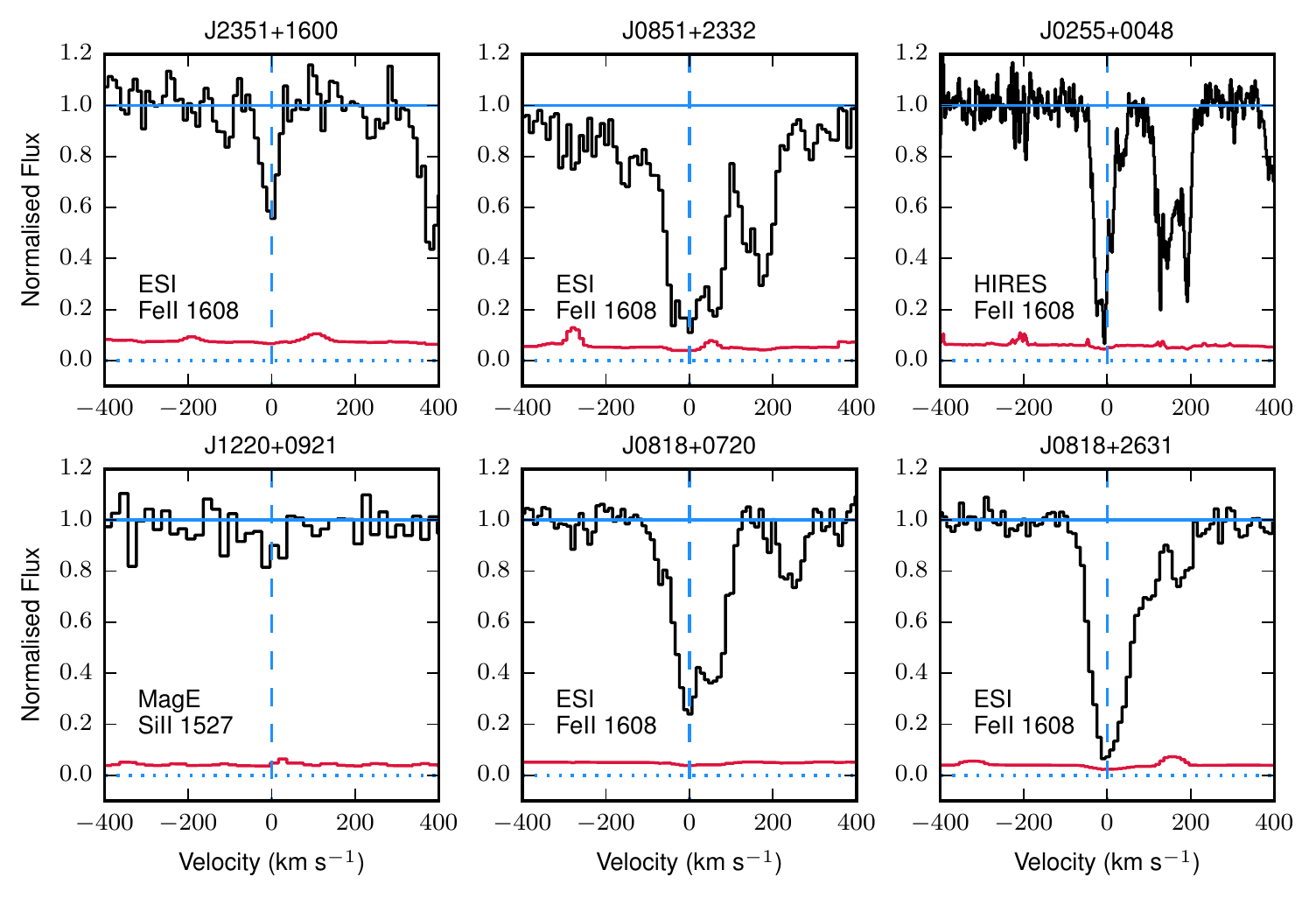}
	\caption{Selected low-ionisation lines from each DLA in our sample. The spectra (black) are continuum normalised, with the 1$\sigma$ flux error shown in red. In each panel the instrument and transition are listed. Lines indicating zero flux (blue dotted line), the normalised continuum level (1.0, blue solid line) and the zero velocity (blue dashed line) are included. Note that many of these lines are saturated and so are not used to derive metallicities.}
	\label{fig:metal_lines}
	\vspace{-1em}
\end{figure*}

\subsection{MUSE observations and data reduction}

MUSE integral field spectroscopy of the six DLA fields was acquired in service mode between June 2015 and April 2016 under ESO programmes 095.A-0051 and 096.A-0022 (PI Fumagalli). Observations were split into sets of 1480s exposures. While four fields were completed with 6$\times$1480s exposures, J0851+2332 and J0818+2631 were only partially completed with 4$\times$1480s and 2$\times$1480s respectively. The observations were taken in dark time with seeing ranging from 0.7 arcsec to 0.9 arcsec using the Nominal Wide Field Mode, with clear conditions. In each sub-exposure the quasar was  centred in the field of view, and 
small dithers combined with
instrument rotations in increments 
of 90$^{\circ}$ were made to improve the quality of the final data product.

The initial reduction of the data is carried out using the ESO MUSE pipeline \citep{MUSEPipeline} (v1.6.2). The pipeline carries out bias, dark and flat field corrections, calibrates the data in wavelength and astrometry, and applies a basic illumination correction. This initial reduction is however limited by the accuracy of the illumination correction and therefore we additionally post-process the data with {\sc CUBExtractor} (Cantalupo, in prep.) to further improve the quality of the illumination correction and sky subtraction (see, e.g. \citealt{Borisova2016,fumagalliLLS,fumagalliUVB}  for details). 

Following this post-processing, the final datacubes are created from the mean of all sub-exposures, and we additionally produce median coadds and datacubes from even and odd numbered sub-exposures to produce independent sets of data for verification processes (see Sect \ref{sec:veto}). The final data product for each field covers a field of view of approximately $1\times 1$ arcmin$^2$, covering $4750-9354$ \AA\ with 1.25 \AA\ binning. Regions of sub-exposures affected by stray-light, near the edges of the cube, are masked before stacking. We also mask 2-5 pixels around the edge of the combined cubes to remove low quality data and very noisy pixels. The field of DLA J0255+0921 further requires substantial masking due to a bright (V$=11.2$) star lying at 46 arcsec from the quasar, resulting in a slightly smaller final field of view. 

As a last step, we calibrate the absolute astrometry and verify the quality of the data against imaging and spectra from the Sloan Digital Sky Survey (SDSS; \citealt{SDSS}). The absolute astrometry of the datacubes is calibrated relative to the quasar position, keeping the relative astrometry within each field as derived with the MUSE pipeline. The spectrophotometric calibration of the datacubes is then checked against SDSS by extracting broadband $r$ and $i$ images from the datacubes and carrying out aperture photometry on brighter stars. Only J0255+0048 is found to require an offset in the form of a
constant multiplicative factor of 1.12 applied to the MUSE data, which brings the flux scale in line with SDSS. All spectra reported in this work have been converted to vacuum wavelength and have had a barycentric correction applied. Comparison to SDSS and high-dispersion spectra of the quasars show excellent agreement. 

\begin{table*}
	\centering
	\begin{tabular}{llccccl}
		\hline
		Field & Instrument  & Exp. Time & S/N           & Resolution    & Wavelength Coverage   & Reference                  \\ \hline
		      &             & (s)       & (per pixel)   & (km s$^{-1}$) & (\AA)               &                           \\ \hline
		J2351+1600  & ESI         & 3600      & 12            & 37            & 3995-10140            & \cite{fumagalli2014}      \\
		J0851+2332 & ESI         & 3600      & 19            & 37            & 3995-10140            & \cite{fumagalli2014}      \\
		J0255+0048  & ESI         & 2400      & 16            & 37            & 3995-10140            & \cite{fumagalli2014}      \\
		      & X-Shooter   & 3320-3600$^a$      & 30            & 40-60$^a$         & 3100-18000            & \cite{Lopez_XQ100}        \\
		      & HIRES       & 20200     & 15            & 6.3           & 5800-8155$^b$             & \cite{Prochaska_HIRES}    \\
		J1220+0921  & MagE        & 3000      & 22            & 71            & 3100-10360            & \cite{Jorgenson_MagE}     \\
		J0818+0720 & ESI         & 3600      & 18            & 56            & 3995-10140            & \cite{fumagalli2014}      \\
		J0818+2631 & ESI         & 3600      & 28            & 56            & 3995-10140            & \cite{fumagalli2014}      \\ \hline
\multicolumn{7}{c}{$^a$Differs between spectograph arms. $^b$With gaps between echelle orders.} 
	\end{tabular}
	\caption{A summary of the spectroscopic data used to establish the absorption properties of the DLAs.}
	\label{tab:spec_data}
\end{table*}

\subsection{Absorption line spectroscopy}

In order to measure the absorption properties of the DLAs, spectra with higher resolution than MUSE (R$\simeq$2000 at 5500\AA) are required, particularly for narrow absorption features and to resolve saturation in strong absorption lines. In this work, we use the same data originally presented in \cite{fumagalli2014}, but we refine the measurement of the \HI\ column density by improving the determination of the quasar continuum. 

The spectroscopic data available are described in Table \ref{tab:spec_data}. For all DLAs in our sample, we have moderate dispersion spectra ($R> 5000$) over the optical range, encompassing the DLA \lya\ and common low ions (e.g. \SiII\ and Fe{\scriptsize~II}), which are used to estimate the DLA metallicity. For the majority of the sample, this is ESI data, while in the case of J1220+0921 the spectrum is from MagE (\citealp{Jorgenson_MagE}). Finally, J0255+0048 has additional higher dispersion data from HIRES (\citealt{Prochaska_HIRES}) covering some low ions at $R\approx 50,000$, and an X-Shooter spectrum \citep[$R\approx 6000$][]{Lopez_XQ100} with higher signal-to-noise ratio (SNR) than the ESI data and coverage of the near infrared.

\subsubsection{\HI\ column densities}

Following the re-analysis in \cite{fumagalli2017} of the \HI\ column density for DLA J0255+0048, we revise 
the original values presented in \cite{fumagalli2014} for the entire sample. 
In this original work, we made use of a local continuum determination, which was later found to underestimate the column density by $0.1-0.2$ dex, due to the fact that the \lya\ absorption lines of the DLAs often coincide with the H$\beta$ and O VI emission lines of the quasars given the relative redshift of the DLAs with respect to the quasars. An example of this can be seen in Fig. \ref{fig:column_densities} where for J0255+0048 (top right) the bump in the revised continuum estimate arises from the H$\beta$ and O VI emission lines of the quasar (which blend together due to the line broadening). In fitting the column densities of DLAs the damping wings provide the primary constraint, hence if the emission lines in the quasar spectrum are not accounted for the column density can be underestimated. This overlap is a consequence of the selection in \cite{fumagalli2014}, where DLAs have similar redshift separations with respect to the quasar in order to exploit the ``double-DLA'' technique. In this work, we refit the \NHI\ column densities replacing the original local continuum determination by the \citet{Telfer} composite quasar spectrum. For each sightline, the template continuum power-law slope and normalisation are adjusted to fit the quasar continuum over the \lya\ forest, and the \HI\ column densities of the DLAs are then estimated by fitting a Voigt profile to the \lya\ transition at the redshift of the DLA. A characteristic uncertainty of $0.1$ dex is assigned to our determinations. Final values of column density are listed in Table \ref{tab:sample}, while a gallery of Voigt profile fits is in Fig. \ref{fig:column_densities}.

During this analysis, we noted that the composite quasar spectrum provides a poor fit to the continuum of quasar J0851+2332, which appears to be significantly lower (approximately a factor of 2) blueward of the DLA \lya. This feature cannot be modelled as a high redshift partial Lyman limit system\footnote{A Lyman Limit System (LLS) is an absorption line with $10^{17.5}$ $<$ (\NHI / cm$^{-2}$) $<$ $10^{20.3}$, such that the system is optically thick to ionising radiation below the Lyman limit (i.e. $\rm \lambda<$912\AA). LLSs are thus characterised by breaks in quasar spectra below the Lyman limit.}, as it would require a redshift above that of the quasar. 
Furthermore, this break is observed in the MUSE, SDSS and ESI spectra, ruling out an artefact with the data. For this reason, in this sightline, we use a manually estimated continuum, found by fitting a spline though points in the forest believed to represent the unabsorbed continuum. Using this model we reach a column density of $\log (N_{\rm HI} / \rm cm^{-2})=21.15$.
We note that, despite some degree of subjectivity in this estimate, 
the column density is mostly insensitive to changes in the continuum, as its upper bound is set by the width of the core of the \lya\ line in this case.
Indeed, if we fit the composite quasar to the spectrum between 5800 and 9000 \AA, we obtain a column density of $\log (N_{\rm HI}/\rm  cm^{-2})=21.25$, only marginally different from our previous estimate. To capture this discrepancy, for this DLA we adopt an uncertainty of $\pm 0.2$ dex.
Finally, when fitting DLA J1220+0921, we note that the \lya\ of the DLA at $z_{\rm DLA}=3.090$ lies close to the Ly$\beta$ of a proximate DLA (pDLA) at $z_{\rm pDLA}=4.1215$. For this line of sight, we therefore also include the Ly$\beta$ transition of this pDLA in our fit. 

\begin{figure*}
	\centering
	\includegraphics[width=\textwidth]{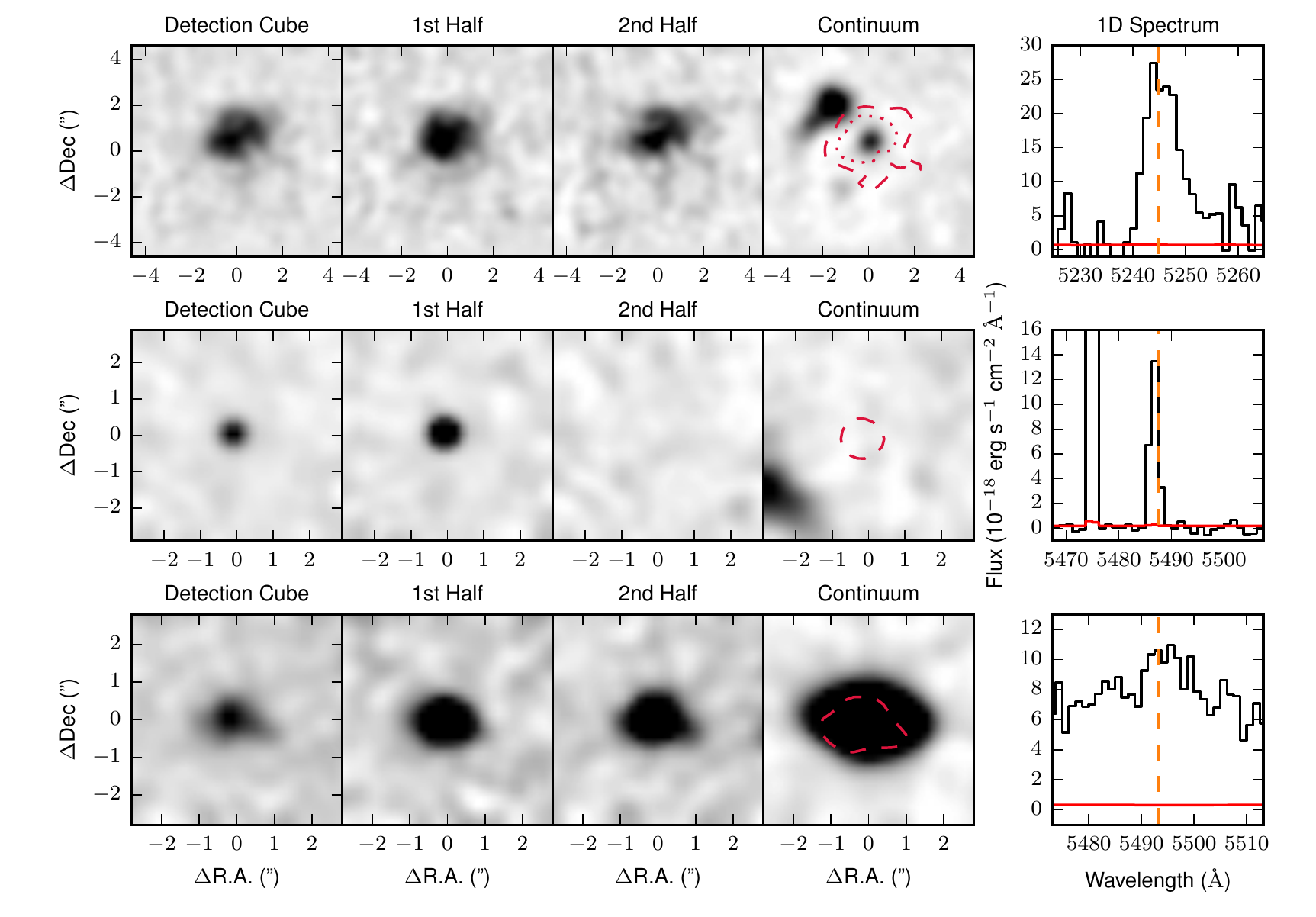}
	\caption{Three examples of LAE candidates that illustrate the three most common occurrences in our identification procedure. The first object (top row) is a high confidence LAE, while the later two are examples of false candidates: a cosmic ray and a low redshift galaxy, respectively. The first three columns in each row show optimally extracted images from the detection cube after continuum subtraction, and the two independent coadds (labelled 1$^{\rm st}$ and $2^{\rm nd}$ half). All three images are shown on a linear scale with the same stretch for a single candidate. The fourth column shows the reconstructed $r$-band image of the same fields with surface brightness contours taken from the detection map of the candidate emission line. All images are smoothed with a Gaussian kernel with full width at half maximum of 3 pixels. The right column shows 1D spectra of the detected emission line for each object (black) with the flux error (red) and the line centre (orange dashed vertical line).}
	\label{fig:veto_example}
	\vspace{-1em}
\end{figure*}

\subsubsection{Metallicities}

In order to calculate the metallicities of the DLAs, we adopt the metal ion column densities compiled in \cite{fumagalli2014}. These values are based on the apparent optical depth method using unsaturated transitions, or are bracketed by upper and lower limits. Table \ref{tab:sample} indicates the ions used to estimate the metallicity of the DLAs in the sample. \cite{fumagalli2017} showed that in the case of J0255+0048 the \SiII\ column density estimated with Voigt profile decomposition of HIRES and X-Shooter data was consistent with the value obtained with ESI and the apparent optical depth method. We therefore conclude the ionic column densities to be robust and do not re-estimate them. 
As the transitions used do not lie in the \lya\ forest, 
the continuum level can be estimated accurately from the data without the need of a quasar template as done for the \HI\ transition. 
Fig. \ref{fig:metal_lines} shows strong low-ionisation absorption lines for each DLA. Some of the lines shown are saturated and were not used to calculate metallicities. From the ionic column densities and the \NHI\ values measured above combined with the solar elemental abundance pattern \citep{Asplund}, we calculate the metallicities of the DLAs, which are listed in Table \ref{tab:sample}.

\section{Search for Galaxy Associations}\label{sec:GalSearch}

To identify galaxy associated to the DLAs, we have conducted a search for both Ly$\alpha$ emitters (LAEs) and Lyman break galaxies (LBGs) by searching for emission line objects in the MUSE datacubes and fitting redshifts for continuum objects detected in deep white-light images reconstructed from the datacubes. 
No candidates were found in the continuum object search to a magnitude limit of $m_{r} < 25$ mag, thus the emission line search has been adopted as our primary method of identifying galaxy associations. The search for continuum objects is nevertheless described for completeness.

\subsection{Search for Ly$\alpha$ emitters}\label{sec:LAEsearch}

We have conducted a search for LAEs with $z_{LAE} \simeq z_{DLA}$ with a velocity window of $\pm1000$ km s$^{-1}$ over the full MUSE field of view. Initially, the mean coadded cubes are trimmed in the wavelength direction to restrict the wavelength range to that of \lya\ over the velocity interval of interest around the DLA of each field, plus a margin of 300 MUSE channels on either side (375 \AA). 

This first slice of the cube retains a sufficient number of channels to perform continuum subtraction of the quasar and other continuum-detected objects in the field using the utilities distributed in {\sc CubExtractor}. The $\pm1000$ km s$^{-1}$ velocity range around the DLAs is masked during this process, to ensure that no emission-line objects are 
subtracted close to the DLA redshift.
Because of this masking, the continuum subtraction in this region is performed by  extrapolating the continuum from the unmasked region. 

After this step, the resulting cubes are more finely trimmed to the wavelength range of interest, plus a margin of one channel to prevent extended objects from being truncated. This continuum-subtracted datacube then becomes the detection cube for our search.
The mean, median and two independent coadded datacubes are also trimmed to the same wavelength range as the detection cube for quality control purposes, as described below. Finally, we run {\sc CubExtractor} over the detection datacubes. The first step in this process is to rescale the data variance in order to match the root-mean-square (RMS) of the flux.
This rescaling is needed to correct for small, albeit significant, degree of correlation in the data following the drizzling process (see Appendix \ref{ap:robust}). 

We then convolve the detection datacubes with a two pixel boxcar spatially, grouping connected pixels above a SNR threshold of 2. These groups are considered a detection if the following criteria are met: i) the object has at least 40 voxels above the threshold; ii) it spans at least 3 wavelength channels at a single spatial position; iii) it has an integrated SNR $\ge 7$. A segmentation map identifies voxels which are above the SNR threshold and are part of a detection.
Segmentation maps are further examined to ensure that connected regions resemble point-like or extended sources (i.e. they are not extremely elongated in a single direction).  

The main source of spurious candidates are cosmic rays and sky line residuals near the edges of the IFUs, which can generate narrow spatial and spectral fluctuations. 
In order to filter some of these artefacts, we extract spectra using the 3D segmentation map produced by {\sc CubExtractor} from the even and odd  datacubes, and we retain only candidates with SNR $\geq 5$ in both coadds. This cut effectively rejects cosmic rays, which appear only in one of the two datacubes. Furthermore, we reject objects that differ significantly in SNR between the two independent coadds (i.e. $\rm \Delta$SNR $> 3$), as these are likely to be associated with sky line residuals or cosmic rays. 

The remaining candidate LAEs are then inspected, using optimally-extracted images (see \citealt{Borisova2016}) from the mean, median, independent coadds and detection datacubes. During this step, we find that comparing the two independent coadds is the strongest discriminant to reject objects which appeared to differ significantly in morphology or position between the images. Finally, the 3D segmentation map is projected onto a 2D grid to extract the object spectrum over the full MUSE wavelength range. Inspection of these spectra enables us to cull other emission lines (commonly $\rm [OII]$) or the residuals of continuum objects, retaining only {\it bona-fidae} LAEs. Bright continuum objects in the field of view can leave residuals during the continuum subtraction, these are often detected by {\sc CubExtractor}. These are unrelated to sky line residuals or cosmic rays and can easily be identified by looking at the datacube without continuum subtraction.
This procedure yields 14 candidate detections over the six DLA fields, as summarised in Table \ref{tab:LAEs}. 

Fig. \ref{fig:veto_example} provides examples for three illustrative cases in our identification procedure. The first example is a high confidence LAE (ID85 from field J1220+0921), where the bright core of the object does not shift in the independent coadds indicating the detection is robust. The second example is a highly-significant detection which, however, only appears in one of the independent coadds, indicating that this candidate is a cosmic ray strike. Examples as significant as this are very rare. The last example is a source that appears strong and marginally extended in the detection cube, however it is much brighter in both independent coadds once the continuum is not subtracted. This feature, combined with the properties of the $r$-band image and 1D spectrum, makes it clear that this candidate is a bright low-redshift galaxy and that the detected feature is in the residual associated with continuum subtraction or an emission line other than \lya.

\begin{figure}
	\centering
	\includegraphics[width=0.5\textwidth]{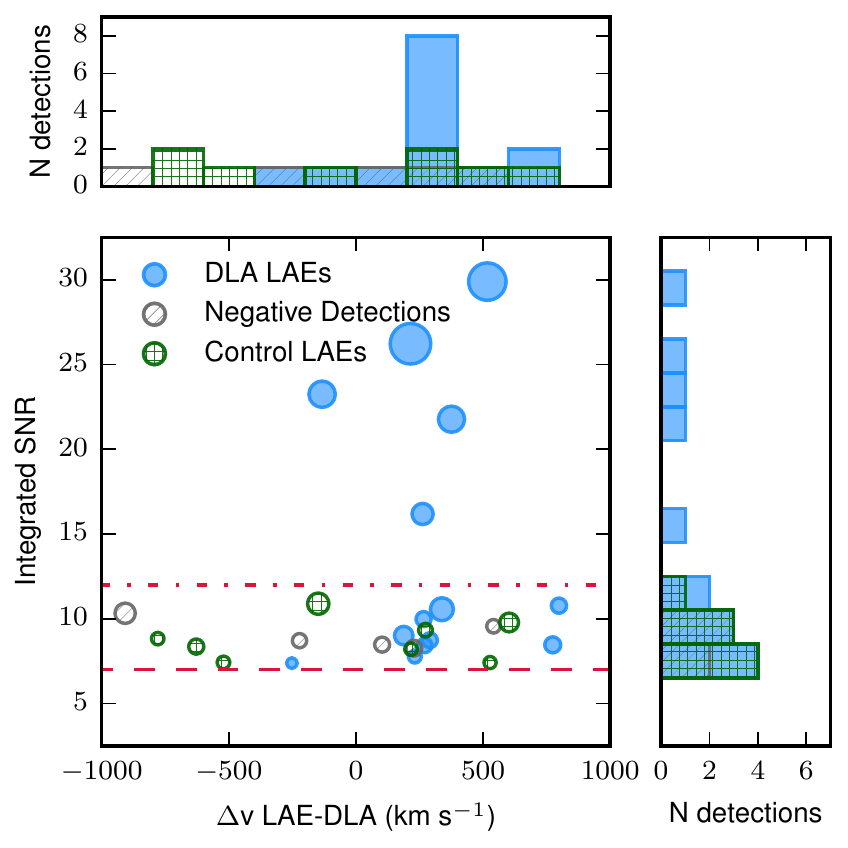}
	\caption{The distribution of velocity separations ($\Delta \rm v$) between DLAs and candidate LAEs against integrated emission line signal-to-noise ratio. Real detections (blue, filled), LAEs selected in control windows (green, gridded) and negative false detections (grey, hatched) are shown. The number of voxels in the segmentation map for each candidate are indicated by the sizes of the symbols. The histogram on each axis shows the distribution of candidates over that single parameter. }
	\label{fig:SNR_vs_deltav}
	\vspace{-1em}
\end{figure}

\subsection{Testing the robustness of LAE identifications}\label{sec:veto}

The resulting candidates from the procedure described above vary in integrated SNR from 7.4 to 29.9. While all detections are robust from a statistical point of view, it is worth examining where an additional cut in SNR is warranted to avoid additional spurious detections that are not rejected in our procedure, especially given that the noise field is non Gaussian. To this end, we have repeated the detection procedure described above, but with the flux values in the datacubes flipped in sign (hereafter the negative datacubes) to explore whether noise fluctuations appear in our selection, and with what SNR. This practice is a standard technique in imaging observations \citep[e.g.][]{Rafelski2009, Hodge2013}.

When searching the negative datacubes, we adopt identical parameters as in the search of real source. However, since real sources with absorption features can generate negative residuals during the continuum subtraction procedure, we remove detections overlapping with bright continuum sources. 
This task generates 5 detections in the negative cubes meeting our requirements. The properties of these detections are compared to the real candidates in Fig. \ref{fig:SNR_vs_deltav} in terms of the velocity offset from the DLA redshift and the detection SNR. 

Additionally, as a further test, we perform the extraction method described in Sect. \ref{sec:GalSearch} over the six datacubes, but this time shuffling the central redshifts (i.e. the DLA redshift) across the fields. 
This experiment yields control source catalogues containing both real LAEs, at a redshift far from the DLAs, and additional spurious detections, if any. The 8 detections produced by this search are displayed in Fig. \ref{fig:SNR_vs_deltav}. This method, unlike the use of the negative detection cubes, does not require symmetry in the noise properties and provides a baseline which we can use to assess if an excess of LAEs clustered to the DLAs is detected in our data.

Two key points are apparent from Fig. \ref{fig:SNR_vs_deltav}. First, the detections from the negative datacubes are all $\rm SNR<11$, while several of the the true detections are at high SNR, up to $\rm SNR=29.9$. Secondly, sources identified in the detection datacubes are significantly more clustered around $\Delta \rm v\simeq+200$ km s$^{-1}$ than both the detections from the negative cubes and the control windows. Due to radiative transfer effects, the Ly$\alpha$ line is expected to be redshifted compared to the DLA redshift by $\approx 100-300~\rm km~s^{-1}$ (e.g \citealt{Steidel10,Rakic11}). 

Therefore, the clustering of sources around $\Delta \rm v\simeq+200$ km s$^{-1}$ compared to both the control sample and the detection in the negative datacubes indicates that our identification procedure yields primarily a sample of true associations.

Our MUSE programme has identified associations with a very high-detection rate, possibly approaching $83\%$ (with at least one detection in 5 out of 6 fields).
However, the fact that a non-negligible number of detections in the negative datacubes pass our selection criteria, we take a conservative approach and establish $\rm SNR=12$ as a threshold for identifying LAEs with high purity, although at the expense of sample completeness. In the following, we refer to these objects (SNR $>12$) as  high-confidence confirmed LAEs, while the remaining sources form a sample of candidate associations that for most part are believed to be real, but for which we cannot exclude the presence of some spurious sources. Deep follow-up observations will be required to determine the nature of each of these sources.

We further note that observations for DLA J0818+2631 suffer quite badly from only having 2 out of 6 exposures completed, meaning that independent coadds contain only a single exposure that presents significant gaps in the reduced datacubes due to the masking around the gaps between the stacks of IFUs in the MUSE FoV. Thus, the search for sources in this field is likely to be incomplete. 

As a last check, we investigate the robustness of our detections, in particular considering whether correlated noise affects the estimates of SNR values and the degree of incompleteness in the detected LAEs. The results of these tests are detailed in Appendix \ref{ap:robust}, where we conclude that correlated noise is not a substantial effect in MUSE data (see also \citealt{BaconUDF}) and that our sample of high-confidence LAEs is highly complete.

\subsection{Identification of continuum objects}

For the identification of continuum detected objects,
we first extract objects using the $r$-band images reconstructed from the cubes running SExtractor \citep{SExtractor} with minimum area of 6 pixels, each above an SNR of 2.0. The SNR threshold was raised for DLA J0818+2631, which had many residuals due to being only partially completed. The resulting segmentation maps are used to define apertures on the datacubes over which we extract spectra for the selected objects.
In this work, we attempt to determine redshifts only for objects with $m_{\rm r} <= 25.0$ mag (corrected for Galactic extinction). This choice is motivated by previous MUSE analyses \citep{fumagalli2015} which have shown how high confidence redshifts for objects fainter than this approximate limit are only obtained in presence of bright emission lines (usually Ly$\alpha$).
As we already search for Ly$\alpha$ emission close to the redshifts of the DLAs, faint LBGs with strong \lya\ emission associated to DLAs will not be missed. 

These spectra are then inspected for emission and absorption lines, as well as characteristic continuum features. Their redshifts are measured by two authors (RM and DJH), either by fitting Gaussian functions to the detected emission lines, or by comparing the 1D spectra with a range of stellar and galaxy templates, including low redshift galaxies and high redshift LBGs. As no $m_{\rm r} <= 25.0$ objects were found to lie close to a DLA in redshift (i.e. within 1000 km s$^{-1}$), these objects are not relevant to our analysis. One LAE detected in the emission line search (id56 in field J0255+0048) has a continuum detection with $m_{\rm r} = 24.58$ but it was not detected in the continuum object search because it was very close to the QSO and was not identified as an independent object. This was not observed with any other $m_{\rm r} <= 25$ objects. The other two LAEs with continuum counterparts are fainter than our magnitude limit for the continuum search and so were not selected. Overall, for $m_{\rm r} <= 25.0$ objects we obtained a redshift completeness of 74\%.

\section{Description of models and simulations} \label{sec:models}

In the following, we compare our observational results to simulations and semi-analytic models to better
understand the constraints they put on the association between DLAs and galaxies. In this section, we provide a detailed description of how different models are produced and analysed. 

The hydrodynamic simulations adopted on this work are taken from the {\sc eagle} suite \citep{EAGLESchaye}, with snapshots post-processed using the {\sc urchin} reverse ray-tracing code \citep{URCHIN} to identify DLAs. In our analysis, we use the post-processed {\sc eagle} snapshots combined with simple prescriptions to populate halos with LAEs to produce mock observations of the correlation between DLAs and galaxies. 
We additionally use a mock catalog based on the \GALICS\ semi-analytic model, designed to produce realistic LAEs \citep{Garel2015}. For this model, DLAs are ``painted'' onto dark matter halos. This simple prescription
allows us to quickly investigate 
the effects of varying the DLA  cross section as a function of halo mass. For a given simulation and prescription we generate a grid which indicates which sightlines through the simulations encounter a painted DLA. These grids are produced by projecting circular kernels centred on halos onto the grid along one axis of the simulation. An additional grid keeps track of position of the DLA in 3 dimensions. Comparisons with the data then allow us to judge to what extent current or future data can distinguish between various models.
This simple model of ``painting'' DLAs onto halos is also applied to the {\sc eagle} simulations, to further gain insight into the properties of DLAs identified with {\sc urchin}.

\subsection{The {\sc eagle} simulations}
Evolution and Assembly of GaLaxies and their Environments ({\sc eagle}, \citealt{EAGLESchaye}) is a suite of cosmological hydrodynamical simulations performed used the {\sc GADGET-3} Tree/SPH code \citep{Springel05} with modifications to the hydrodynamics solver described by \cite{Schaller15}. The simulations incorporate the dominant cooling and heating processes of gas in the presence of the uniform but time-varying UV/X-ray background of \cite{HaardtMadau} as described by \cite{Wiersma09a}. Physics below the resolution scale, such as star and black hole formation and their feedback effects, are incorporated as \lq subgrid modules\rq\ with parameters calibrated against observations at redshift $z=0$ of the galaxy stellar mass function, the relation between galaxy mass and size, and the relation between galaxy mass and black hole mass \citep{EAGLECrain}. The simulations reproduce a number of observations that were not part of the calibration, including the colour-magnitude diagram \citep{Trayford17}, the small-scale clustering of galaxies at $z=0$ \citep{Artale17}, the evolution of the galaxy stellar mass function \citep{Furlong15}, the connection between Active Galactic Nuclei and star formation \citep{McAlpine2017} and of galaxy sizes \citep{Furlong17}, and the evolution of the \ion{H}{I} and ${\rm H}_2$ contents of galaxies \citep{Crain17, Lagos15}. At $z\simeq3$ \eagle\ has been shown to match the observed star formation rate density well \citep{Katsianis2017}, and bears reasonable agreement with the metallicity distributions of high column density absorption line systems \citep{Rahmati2018}.

The {\sc eagle} simulations are performed in cubic periodic volumes, and the
linear extent ($L$) of the simulation volume and number of simulation particles is varied to allow for numerical convergence tests. In this work,  we use simulations L0100N1504 and L0025N0752 from table~1 of \cite{EAGLESchaye}. Briefly these have $L=100$~co-moving megaparsecs (cMpc) and $L=25$ ~cMpc, and a SPH particle masses of $1.8\times 10^6$ ${\rm M}_\odot$ and $0.2\times 10^6$ ${\rm M}_\odot$, respectively; the Plummer equivalent co-moving gravitational softening is 2.66 and 1.33~kpc, limited to a maximum physical softening of 0.7 and 0.35~kpc, respectively. The simulations assume the cosmological parameters of \cite{Planck2014}, and the minor differences with the \cite{Planck2015} parameters adopted in the current paper are unlikely to be important.

To compare to the data presented earlier, we need to identify both DLAs and LAEs in {\sc eagle}. Since neither neutral hydrogen nor \lya\ radiative transfer are directly incorporated in {\sc eagle}, we compute these quantities in post-processing as explained next.

\begin{figure*}
	\centering
	\includegraphics[width=\textwidth]{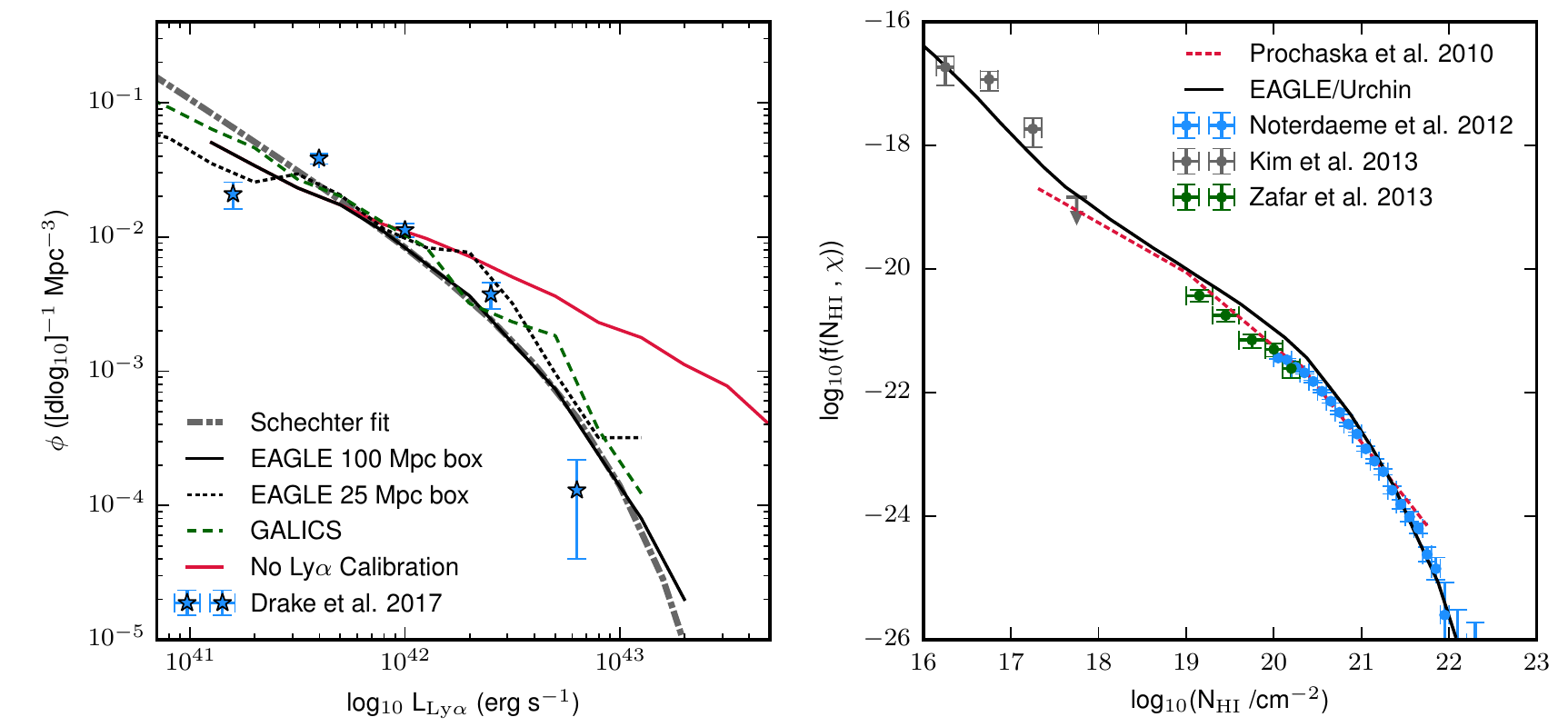}
	\caption{\textit{Left: } The \lya\ luminosity function from the {\sc eagle} LAE sample compared against recent observations (\citealp{Drake2017}, blue stars) and a Schechter function fit to these observations (gray dot-dashed). The calibrated (black solid line) and original (red solid line) luminosity functions for the 100 Mpc box simulation are shown, together with the rescaled luminosity function from the 25 Mpc {\sc eagle} box (dotted line). The luminosity function from \GALICS\ is plotted (green, dashed), and shows excellent agreement with the data without need for calibration.
		\textit{Right:} The column density distribution function derived from the {\sc eagle} simulation after post-processing with the {\sc urchin} radiative transfer code, compared against observations at comparable redshift.}
	\label{fig:CDDF}
	\vspace{-1em}
\end{figure*}

\subsubsection{Identifying DLAs in the {\sc eagle} simulations} 
The ionising background in {\sc eagle} is implemented in the optically-thin limit. Self-shielding of gas, allowing for the appearance of DLAs, is computed in post-processing using the {\sc urchin} radiative transfer code described by \cite{URCHIN}. Briefly, this algorithm allows each gas particle to estimate the local ionising intensity it is subject to by sampling the radiation field in 12 directions, with neutral gas in neighbouring gas particles potentially decreasing the local photo-ionisation rate below the \cite{HaardtMadau} optically-thin value. Assuming the neutral fraction of a particle is set by the balance between photo- and collisional ionisation versus recombinations, a reduced ionisation rate increases the particle's neutral fraction, which in turn affects the ionisation rate determined by the particle's neighbours. The impact of a change in the neutral fraction on the photo-ionisation rate and vice versa is iterated until the neutral fraction of each particle converges from one iteration to the next. This post-process step thus yields the neutral hydrogen fraction $x_\ion{H}{I}\equiv n_\ion{H}{I}/n_{\rm H}$ of each SPH particle.
Further details on how radiative transfer affects the neutral fraction as a function of column density and physical processes (e.g. the relative importance of collisional ionisation and photoionsation), can be found in the literature \citep[e.g.][]{Fumagalli2011,Rahmati2013}.

The resulting \HI\ volume density is then projected onto a 8192$^2$ grid along the coordinate $z$-axis of the simulation box, using the Gaussian smoothing described by \cite{URCHIN}.
Applying a column density threshold of $\log(N_\ion{H}{I}/ {\rm cm^{-2}})=20.3$ allows us to identify DLAs\footnote{The redshift path through the simulation box is so small that the contribution of chance alignments of two high-column density systems to the DLA cross section is negligible.}. We also calculate
the $x_\ion{H}{I}$-weighted $z$-coordinate and velocity of particles along each DLA line of sight to obtain the 3D position of each DLA (two spatial coordinates $x$ and $y$, and a redshift coordinate $z$).
The redshift of each DLA allows us to compare DLAs to galaxies in redshift space, although we note that observed DLA redshifts are derived from low-ionisation metal lines rather than \HI\ directly, yet these elements are believed to trace the same phase of the gas.

To compare to the {\sc muse} observations, we analyse the L0025N0752 {\sc eagle} snapshot at $z=3.027$ which is closest in redshift to the data. 
Similarly to the results of \citet{Altay11} based on the 
{\sc owls} simulations \citep{OWLSSchaye}, the column density distribution function
(CDDF, the number density of absorbers per unit column density, per unit absorption distance,
$f(N_{\ion{H}{I}})$) of the post-processed snapshots using {\sc urchin} is in good agreement with observations (Fig. \ref{fig:CDDF}). The data in this figure combines the low column density data ($\log (N_{\rm HI}/\rm cm^{-2})<18$) compiled by \citealp{Kim2013} ($z=2.4 - 3.2$), the multiple power-law fit to the Lyman-limit  and DLA column density range derived by \citet{ProchaskaLLS} at $z=3.7$, and the sub-DLA and DLA data from \cite{Noterdaeme} ($\langle z \rangle =2.5$). At these redshifts, we find that the simulated and observed incidence of DLAs is very similar. The measurements of \cite{Zafar} in the sub-DLA range are also shown, these constraints cover a much broader redshift distribution than the MUSE DLAs ($1.5<z<5$) but show close agreement with the fit of \citet{ProchaskaLLS}.

\subsubsection{Identifying LAEs in {\sc eagle}} \label{sec:EAGLE_LAEs}
Gravitationally bound substructures in {\sc eagle} are identified combining the friends-of-friends, \citep{Davis85} and {\sc subfind} \citep{Springel05, Dolag09} algorithms. 
Physical properties of these \lq galaxies\rq\, such as their centre of mass position and velocity, stellar mass and star formation rate, are computed and stored in a database \citep{EAGLEdatabase}. In {\sc eagle} the star-formation rate of a gas particle is a function of its pressure, it is zero below a metallicity-dependent density threshold and above a pressure-dependent temperature threshold \citep{EAGLESchaye}. SFRs and stellar masses of subhalos are computed by summing over the gas and star particles (respectively) within a given subhalo identified with {\sc subfind}. 
Lacking sufficient resolution to resolve the ISM, we cannot predict from first principles the Ly$\alpha$ properties of the simulated galaxies. We resort instead to a simpler empirically-motivated model.
We populate the simulation with LAEs by selecting simulated galaxies by star formation rate (SFR) and associating a \lya\ luminosity ($ \rm L_{Ly \alpha }$) using the conversion from Eq.~\ref{eq:lyalum} \citep{Furlanetto}. This relation combines hydrogen case B recombination with a standard SFR calibration \citep{Kennicutt1998}. 

\begin{equation}
\qquad \qquad \rm  L_{Ly \alpha } = \frac{SFR}{M _\odot yr^{-1}} \times 10^{42} ~erg ~s^{-1}
\label{eq:lyalum}
\end{equation}

This prescription neglects diffuse emission from the low density intergalactic and circumgalactic medium, but captures the bulk of the emission associated with star formation inferred from stellar population synthesis models under the assumption that $\approx 2/3$ of the recombinations occurring in \ion{H}{II} regions produces a Ly$\alpha$ photon. This is a reasonable approximation for hydrogen, yet may nevertheless yield a large overestimate of the Ly$\alpha$ luminosity of a galaxy because a significant fraction of such photons do not escape the galaxy, due to to scattering and dust (but see below for how we correct this effect). 

The \lya\ luminosity function of LAEs is shown in the left panel of Fig.~\ref{fig:CDDF},
comparing results from L0025N0752 and L0100N1504 with the Schechter fit to the observed luminosity function from \cite{Drake2017}, using the redshift bin $2.92 \le z < 4.00$. As anticipated,
by neglecting dust absorption in \eagle\ the bright end of the luminosity function (starting from $\rm L_{Ly \alpha } > 5 \times 10^{41} erg~s^{-1}$) significantly overpredicts the observed number density of emitters, especially at high SFRs, $\dot M_\star\ge 10{\rm M}_\odot~{\rm yr}^{-1}$. This trend is well documented observationally: star formation rates inferred from the UV compared to those based on Eq.~(\ref{eq:lyalum})
are discrepant if no correction is made for dust (e.g. \citealp{Dijkstra10,Whitaker2017}). Moreover, \cite{Shapley2003} show that roughly a third of the LBG population is not detected in Ly$\alpha$ at all. Also, \cite{Matthee2016} show that at $z\simeq2.2$ galaxies with higher H$\alpha$-inferred SFRs have lower \lya\ escape fractions, thus implying that a correction is required at the bright-end of the simulated luminosity function.

We therefore account for these unresolved physical processes (e.g. dust extinction and escape fraction) by introducing an effective  \lya\  escape correction to the simulated Ly$\alpha$ fluxes. We do so by sub-sampling the simulated star-forming galaxies until they match the Schechter fit to the luminosity function from \cite{Drake2017}. This is done by dividing the Drake et al. Schechter fit by the measured {\sc eagle} luminosity function, and applying this ratio as  probability that a galaxy of a given luminosity will be added to the LAE catalogue. The result of this re-scaling is shown in Fig. \ref{fig:CDDF}, revealing that the sub-sampling performs well for $\rm L_{Ly \alpha } > 5 \times 10^{41} erg~s^{-1}$. Given that the high-confidence MUSE sample extend only down to $\rm 10^{42} erg~s^{-1}$, the {\sc eagle} model appears excellent for comparison to the observations. This resampling technique could be thought of in terms of a  \lya\  duty-cycle, with the time a galaxy spends in an LAE phase being a function of the SFR \citep{Nagamine2010}.

As consistency check, we compute the clustering of sub-sampled LAEs identified in the L0100N1504 {\sc eagle} simulation (Fig. \ref{fig:clustering}). Ideally our identified LAEs should match observational measurements of the clustering amplitude, as typically quantified by fitting a simple power-law model to the correlation function,

\begin{equation}
\qquad \qquad \qquad \qquad \qquad  \rm  \xi(r) = (r/r_0) ^{\gamma}\,.
\label{eq:r_0}
\end{equation}

Observational estimates of $\rm r_0$ for LAEs at this redshift range between 2-4 Mpc (\citealp{MUSEWide}), with $\gamma = -1.8$ typically assumed (e.g. \citealt{Bielby2016,Gawiser07}). \lya\ luminosity typically varies from study to study, and so the samples are not necessarily the same. Recently \cite{MUSEWide} obtained $\rm r_0=2.9_{-1.1}^{+1.0}$ with MUSE, while a narrow-band survey measured $\rm r_0=4.58_{-0.43}^{+0.44}$ \citep{Khostovan18}. As discussed in section \ref{sec:veto}, our high-confidence sources with SNR>12 have luminosity $ \rm  L_{Ly \alpha } > 10^{42}$ erg s$^{-1}$. Based on this selection, we draw comparison to the study of \cite{Gawiser07}, where a comparable luminosity limit was adopted yielding a value of $\rm r_0=3.6_{-1.0}^{+0.8}$ Mpc. Additionally, this estimate represents a compromise between the highly varying literature values. Fig. \ref{fig:clustering} shows that the clustering of our selected LAEs in {\sc eagle} is higher than $\rm r_0=3.6$ Mpc, favoring instead a value of $\rm r_0=5.1$~Mpc (between $1-15$~Mpc). 
Although the measurement of \cite{Gawiser07} will suffer from limited volume our simulated LAEs have a higher $\rm r_0$ than most observations, we will consider the impact of this offset in Sect. \ref{sec:disc_mod}. Cross-correlations will be less affected by a higher clustering amplitude of one sample than the auto-correlations shown in Fig. \ref{fig:clustering}. 

Together, these comparisons demonstrate our simple model for assigning a \lya\ luminosity to {\sc eagle} galaxies, which is calibrated to reproduce the abundances of LAEs. Although there is tension in the clustering amplitude we conclude in  Sect. \ref{sec:disc_mod} that the agreement is sufficient to enable a valid comparison between the simulations and the data in the current paper. In the following section, we further show how an independent semi-analytic model constructed to reproduce the luminosity function of LAEs agrees well with the prediction derived from the \eagle\ simulations. This semi-analytic modeling represents a cross-check of our results and includes a more physical model of \lya\ escape, independently of our method of modeling LAEs in  {\sc eagle}.

\subsection{The \GALICS\ semi-analytic model} \label{sec:mock}

As an alternative model to the one based on \eagle, we use mock catalogs of LAEs based on the model of \cite{Garel2015} which combines the \GALICS\ semi-analytic model with numerical simulations of \lya\ radiation transfer in galactic outflows \citep{Verhamme2006, Schaerer2011a} to predict the observed \lya\ luminosities of galaxies. As shown in \cite{Garel2015, Garel2016}, the model can reproduce various statistical constraints on galaxies at high redshift, such as the abundances of LAEs at $3<z<6$. The mock lightcones used in this study were extracted from the \GALICS\ cosmological simulation volume (L$_{\rm box}$ = 100~h$^{-1}$ cMpc) and were specifically designed to match the redshift range, geometry and depth of typical \lya\ surveys with MUSE (see \citealt{Garel2016} for more details about \GALICS\ and the mocks). We refer to these mock catalogues hereafter as \GALICS, including the additional radiative transfer.

\begin{figure}
 	\centering
 	\includegraphics[width=0.5\textwidth]{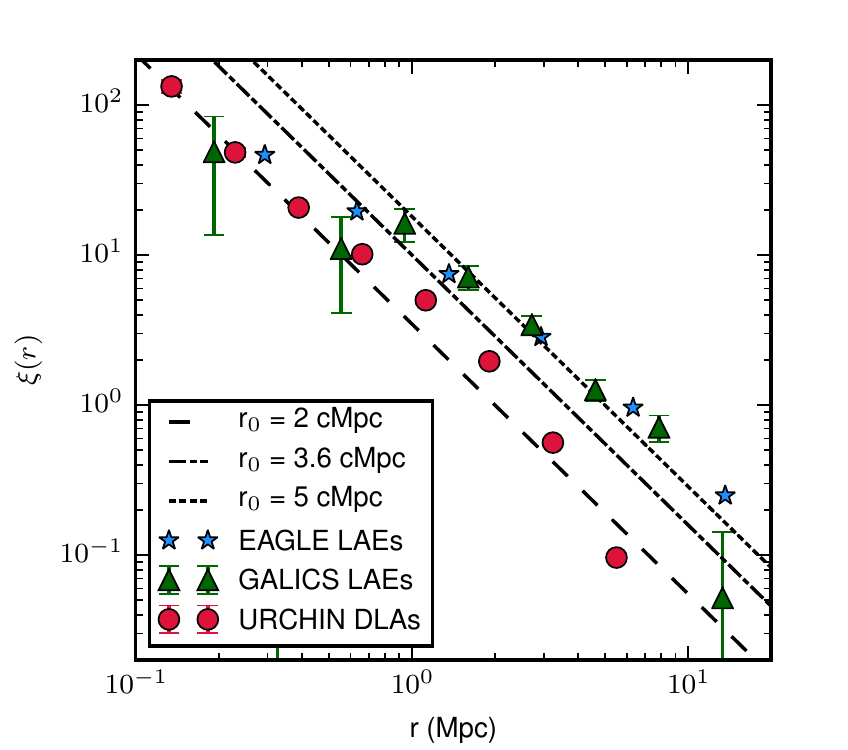}
    \caption{The two-point correlation function of LAEs ($\rm L_{Ly \alpha} \ge 10^{42}$ erg s$^{-1}$, {\em blue stars:} sub-sampled LAEs identified in the L0100N1504 \eagle\ simulation, {\em green triangles}: LAEs from the \GALICS\ SAM) and of DLAs identified in the L0025N0768 \eagle\ model ({\em red circles}).  Three example power-law functions, $\xi(r)=(r/r_0)^\gamma$, with $\gamma=-1.8$ are shown for different values of $r_0$ as per the legend.}
 	\label{fig:clustering}
 	\vspace{-1em}
 \end{figure}

Fig. \ref{fig:CDDF} (left) shows the predicted \lya\ luminosity function from \GALICS, compared to MUSE deep field observations \citep{Drake2017}. Fig. \ref{fig:clustering} also demonstrates that the clustering of the LAEs in the lightcone is in close agreement with that of \eagle. Thus, the \GALICS\ mock catalog represents an excellent way to cross-check predictions from {\sc eagle}, and further test our selection of LAE-like galaxies from the simulation. The LAE mock, however, does not simulate neutral hydrogen, and we describe next a simple model which can be applied to both \GALICS\ and {\sc eagle} (as alternative to the {\sc urchin} post-processing) to populate dark-matter halos with DLAs.

\subsection{A halo prescription for DLAs}

\begin{figure*}
	\includegraphics[width=\textwidth]{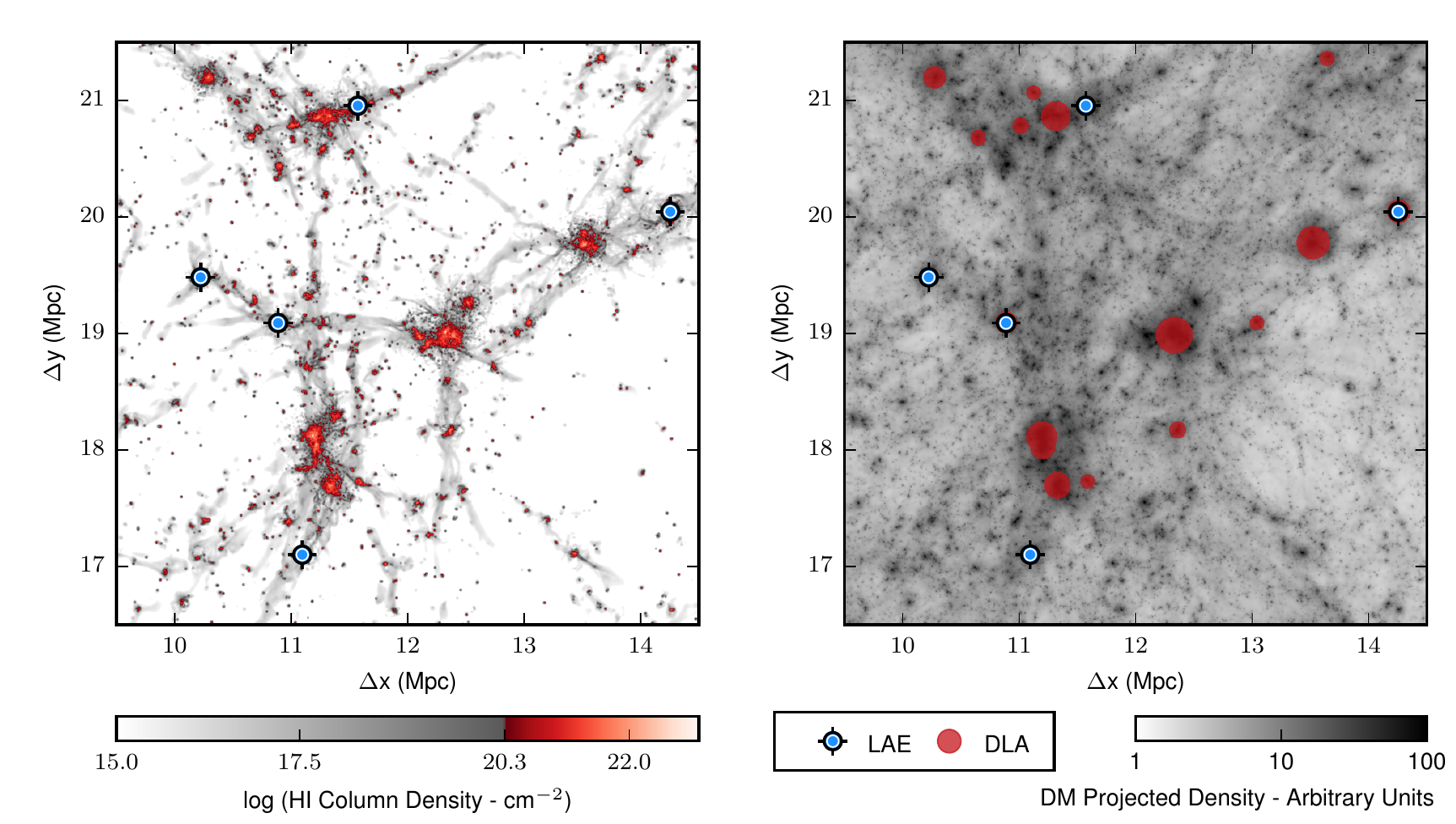}
	\caption{A comparison of the two DLA modelling techniques utilised in this paper. Each panel shows the distribution of DLAs and LAEs (with $\rm L_{Ly \alpha} \ge 10^{42}$ erg s$^{-1}$) from the same \eagle\ 25 Mpc simulation, shown on the same physical scale. Both plots are projected over the full 25 Mpc box length. \textbf{Left:} The DLAs have been computed from the hydrodynamics of the \eagle\ simulation using the {\sc urchin} post-processing. The color-map corresponds to \NHI\ with sightlines above the DLA column density threshold (log(\NHI / cm$^{-2}$) $\geq$ 20.3) highlighted in red.
		\textbf{Right:} A demonstration of the halo painting scheme used to populate the \eagle\ simulation with DLAs based on the halo catalogue. Each halo above the mass threshold is painted with a circular cross-section that is proportional to its halo mass for $\alpha=0.75$. The projected dark matter density is plotted for reference in greyscale, in arbitrary units, using Py-SPHViewer \citep{alejandro_benitez_llambay_2015_21703}.}
	\label{fig:sim_compare}
\end{figure*}

As well as exploiting the hydrodynamics of the {\sc eagle} simulations to predict the position and properties of DLAs, we employ a second model by assigning DLAs to the dark matter halos from the simulations via a simple halo ``painting'' model. This is similar to the model put forward in \cite{FontRibera12}, and updated in \citealp{BOSS2018}, in which DLA cross sections are assigned to halos as a function of halo mass. With this model, it is possible to quickly adjust the parameters to match observations,  such as the large scale clustering of DLAs as measured in BOSS. 

In this model, the relation between DLA cross section and halo mass is described by Eq.~\ref{eq:halomod}, where $\rm \Sigma (M_{\rm h})$ is the DLA cross-section for a halo of mass $\rm M_h$ above some minimum halo mass $\rm M_{min}$ with a power-law slope of $\alpha$ and a zero-point $ \rm \Sigma_0$.
\begin{equation}
\qquad \Sigma (\rm M_h) = \Sigma_0 (M_h/10^{10} M_{\odot})^{\alpha}   \qquad (M_h > M_{min})
\label{eq:halomod}
\end{equation}

The halo catalogues used in this work are obtained from the \eagle\ database \citep{EAGLEdatabase} Friends-of-Friends table for the 25 Mpc and 100 Mpc \eagle\ boxes.
While the 100 Mpc box suffers from resolution effects at low halo masses ($M_{200} \simeq  10^9$\msun), the higher-resolution 25 Mpc box suffers from a limited volume that contains few massive halos (above $M_{200} =  10^{13}$\msun). For these reasons, we combine both the 25 and 100 Mpc simulations, whereby the higher resolution simulation allows us to study halo model parameters which extend the cross-sections to low masses, while the larger box provides better convergence at high halo masses and on larger scales (see also \citealt{Pontzen2008}). 

To populate halos with the cross sections specified in Eq. \ref{eq:halomod}, we generate a 8192$^2$ grid along the $z$-axis of the simulation with circular kernels representing DLAs centred on halos, the size of which varies with halo mass. While values of $\rm M_{min}$ and $\alpha$ are fixed to those from \cite{FontRibera12} to reproduce the observed large scale clustering of DLAs, $\rm \Sigma_0$ is fit for each value of $\rm M_{min}$ and $\alpha$ such that the umber of DLAs per unit path length ($\ell_{DLA}(\rm X)$) in the 100 Mpc simulation matches  $\ell_{DLA}(\rm X)$ of the {\sc urchin} DLAs in the 25 Mpc box. Hence, we calibrate the cross section to $\ell_{DLA}(\rm X)=0.0948$, which is in good agreement with observational estimates at $z\simeq3.5$ \citep[e.g.][]{XQ100_lX}. We then transfer the same calibration onto the other simulations. In order to obtain 3D coordinates for the DLAs, 
we use the location of the DLA parent halo, or take an average in the case where DLAs overlap. The periodic boundary of the box is taken into account during the projection. 

A powerful feature of this model is the possibility to quickly explore how different parameters impact the small-scale clustering of galaxies around DLAs, which is the quantity probed by our observations. For this reason, we implement different values for the model parameters, as summarised in Table \ref{tab:BOSS_param}. As this simple scheme is independent of the hydrodynamics of a simulation, we have also applied it to the \GALICS\ mock catalog described in Sect. \ref{sec:mock}.

Fig. \ref{fig:sim_compare} shows the halo painting model applied to a section of the EAGLE 25 Mpc simulation for the $\alpha=0.75$ model (right) alongside the \NHI\ column density map from the \eagle/{\sc urchin} post-processing. Qualitatively, this choice of parameters produces a covering factors of DLAs which closely resembles the result of the simulation, with the difference of a sharp cut-off at a fixed minimum mass, which does not apply to the \eagle\ simulations. 

Moreover, it is also apparent in Fig. \ref{fig:sim_compare} (left) that the \eagle\ simulations contain small clumps of \HI, some of which reach DLA column density. Being close to the resolution limit of the simulations, we cannot assess whether the \HI\ properties of these halos are fully converged and physically meaningful. Although small in cross-section, the clumps are numerous, and often far from massive galaxies. It may be these clumps which suppress the clustering of the {\sc urchin} DLAs with respect to the results from the halo-painting model (Fig.~\ref{fig:clustering}).

\begin{table}
	\renewcommand*{\arraystretch}{1.4}
	\centering
	\begin{tabular}{llll} 
		\hline
		$\alpha$ & $\rm M_{min}$ (\msun\ ) & $\rm M_{max}$ (\msun\ ) & $\rm \Sigma_0$ (Mpc$^2$)  \\
		\hline
		1.1  & 3.0$\times$10$^{9}$  & 1.5$\times$10$^{13}$  & 0.000678 \\
		0.75 & 5.9$\times$10$^{10}$  & 1.5$\times$10$^{13}$  & 0.00327 \\
		0.0  & 2.2$\times$10$^{11}$  & 1.5$\times$10$^{13}$  & 0.123 \\
		\hline
	\end{tabular}
	\caption{Different parameters of the DLA halo model by \protect\cite{FontRibera12} which we adopt in this work.}
	\label{tab:BOSS_param}
\end{table}

\begin{figure*}
	\centering
	\includegraphics[width=1.0\textwidth]{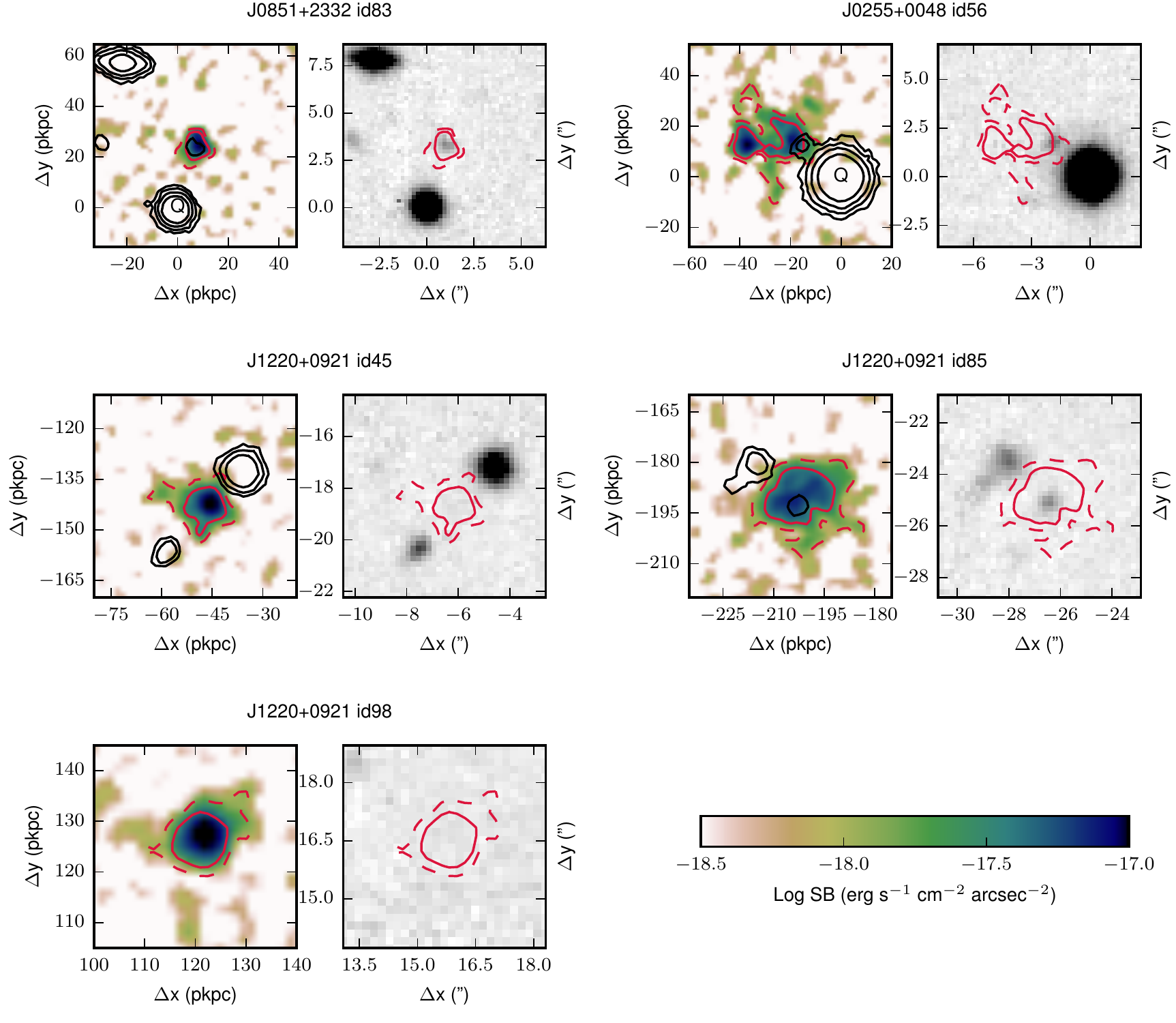}
	\caption{Postage stamps of the high confidence DLA associations. Each object has an optimally extracted \lya\ surface brightness map (left) and an $r$ band image (right). Both images have \lya\ SB contours (red), drawn at 10$^{-18}$ erg s$^{-1}$ cm$^{-2}$ arcsec$^{-2}$ (dashed lines) and 10$^{-17.5}$ erg s$^{-1}$ cm$^{-2}$ arcsec$^{-2}$ (solid lines) . The \lya\ SB map is smoothed with a Gaussian kernel ($\sigma=1$ pixel) and has $r$ band continuum contours overlaid in black. The scale of the SB map is in physical kpc from the DLA, while the $r$ band image is shown in arcsec from the quasar. The quasar position (labeled 'Q') is indicated where visible.}
	\label{fig:lya_SB}
	\vspace{-1em}
\end{figure*}

\begin{figure*}
	\centering
	\includegraphics[width=1.0\textwidth]{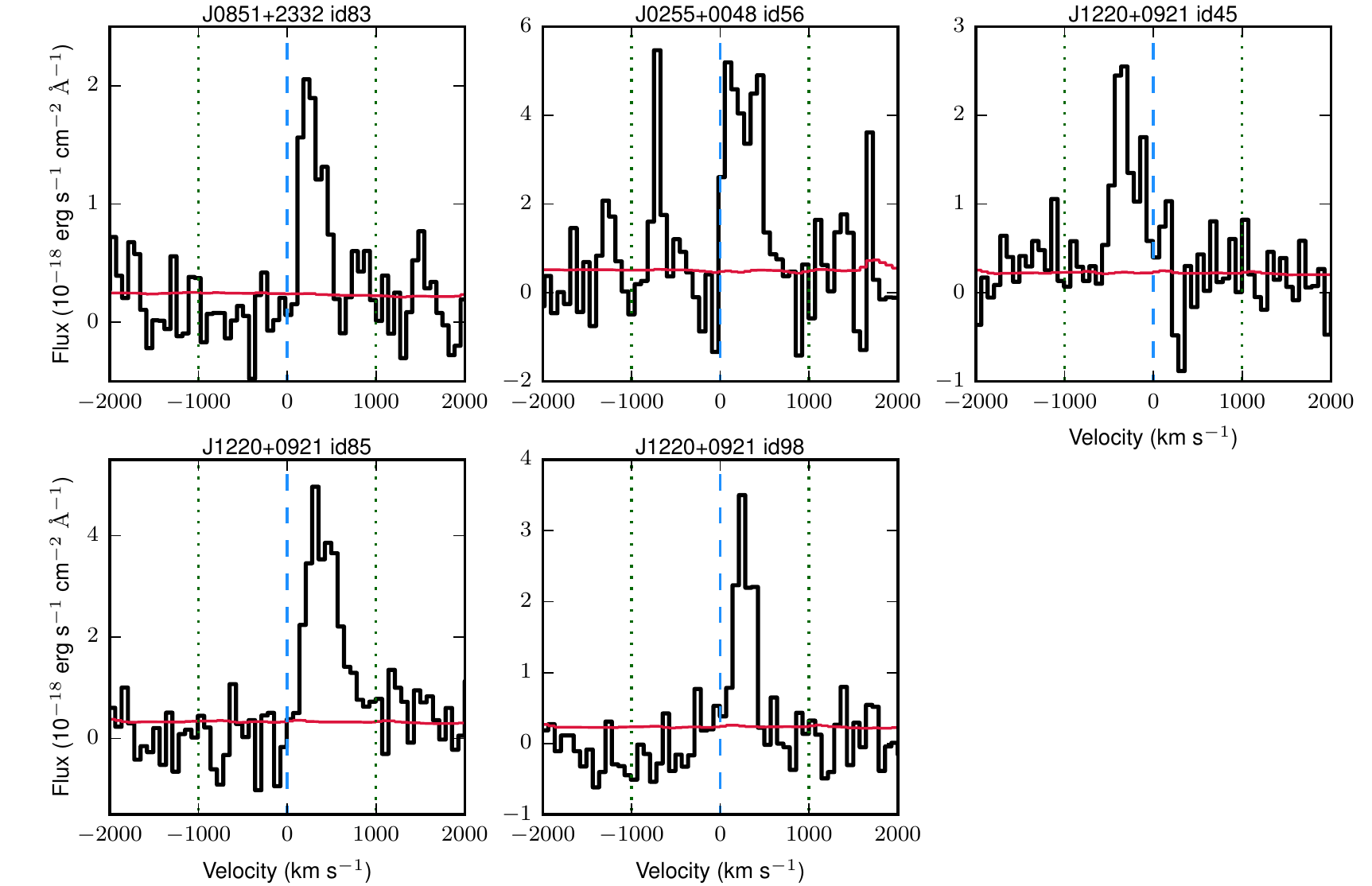}
	\caption{The \lya\ emission lines for the high-confidence DLA associations. The 1D extracted spectrum is shown in black, with the 1$\sigma$ error in red, plotted as a function of velocity offset from the DLA absorption redshift. Vertical blue dashed lines indicate the DLA redshift, while green dotted lines mark the velocity window over which LAEs were extracted ($\pm$1000 km s$^{-1}$). Fluxes have been corrected for Galactic extinction.}
	\label{fig:lya_profiles}
	\vspace{-1em}
\end{figure*}

\begin{figure*}
	\includegraphics[width=\textwidth]{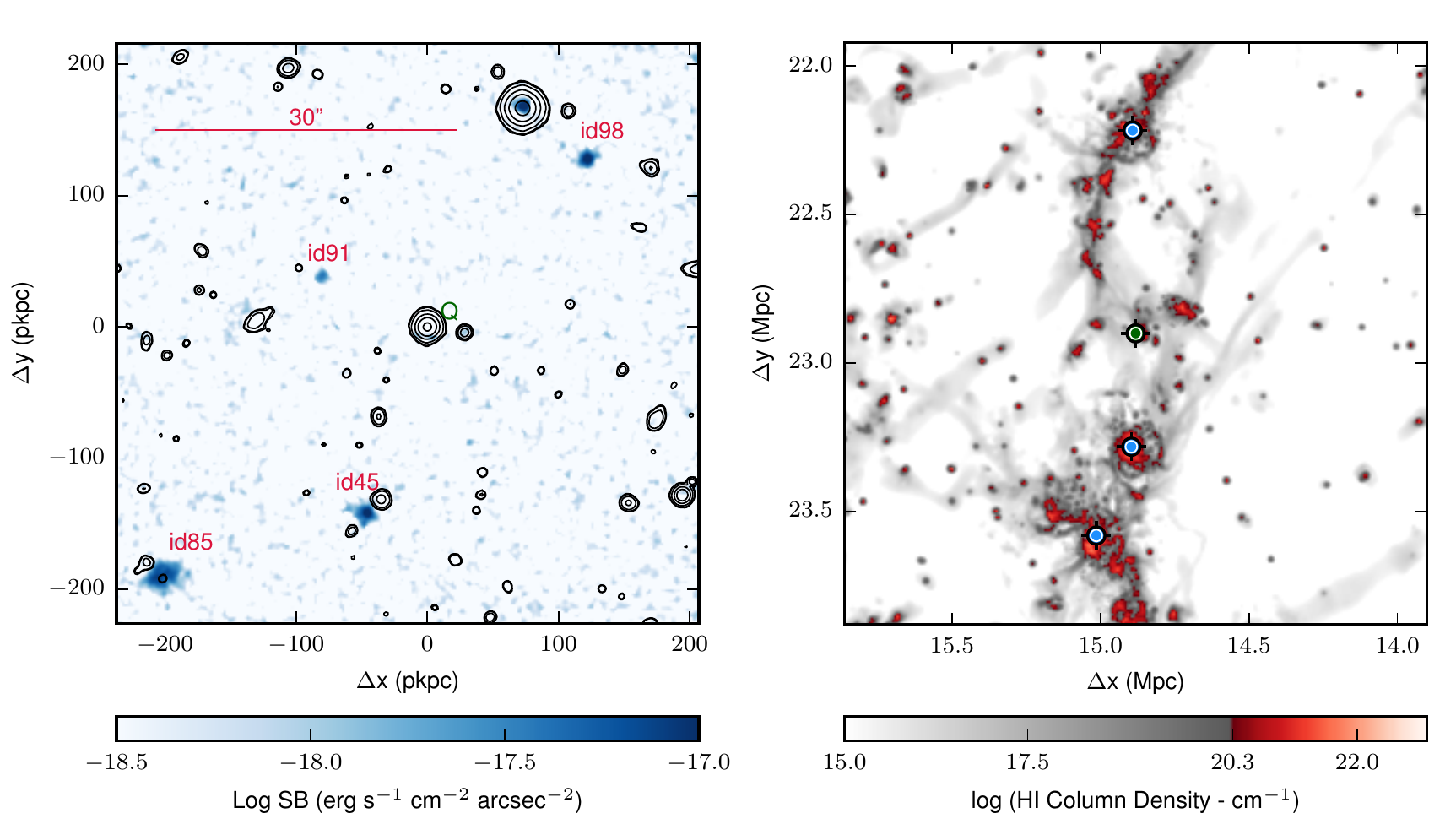}
	\caption{\textbf{Left:} A map of the MUSE field around the DLA J1220+0921 sightline, showing an optimally extracted \lya\ surface brightness map of all four objects detected around the DLA. The pixels inside the segmentation map for each detection are projected onto an image and three adjacent channels of noise is added. The angular scale is shown with the quasar position and the LAE ids (red). $r$-band contours are shown in black for context. The large object to left of id98 with apparent emission is a bright star, the false line emission is in fact a continuum subtraction residual.
		\textbf{Right:} An analogue of the J1220+0921 system extracted from the \eagle\ simulation for illustrative purposes. The \HI\ column density projected though the 25 Mpc simulation is shown in greyscale, with the transition to the red colour-map indicating the DLA column density threshold. The FoV matches that of the MUSE DLA observations in co-moving distance. Also marked are LAEs with $L_{Ly \alpha} \ge 10^{42}$ erg s$^{-1}$ (blue) and the central DLA (green).}
	\label{fig:map_19H7}
\end{figure*}

\section{Properties of the high-confidence associations}\label{sec:results}

\subsection{Notes on individual fields}

Following the search of LAEs and redshifting of continuum-detected sources, we identify five high-confidence LAE associations (with  sufficiently high SNR to be at very high purity) in three out of six DLA fields (J0851+2332, J1220+0921 and J0255+0048, which was previously published in \citealt{fumagalli2017}). 
The derived properties of the detected objects are summarised in Table \ref{tab:LAEs}, 
with both Ly$\alpha$ and $r$-band images shown in Fig. \ref{fig:lya_SB}. Spectra of the Ly$\alpha$ lines are also shown in Fig. \ref{fig:lya_profiles}. In addition to the high-confidence associations, we further identify 9 LAE candidates across five fields, which are shown in Fig. \ref{fig:other_cands} in Appendix \ref{ap:LBGs}. These are detections at lower SNR and likely include a high fraction of true associations, but for which we cannot guarantee the sample purity.
In the following, we provide a brief discussion of the key features of the high-confidence associations.

\begin{table*}
	\begin{tabular}{lllccccccc} 
		\hline
		Id & RA $^a$ & Dec $^a$  & $b$ $^a$  & Ly$\alpha$ Flux $^b$ & m$_{\rm r}$ $^c$ & L$_{\rm Ly\alpha}$ $^b$ & $\Delta v$ $^a$  & D$_{\rm LOS}$  $^a$ & SFR$_{UV}$ $^d$ \\
		& (hrs)  & (deg)  & (arcsec) & (10$^{-18}$ erg s$^{-1}$ cm$^{-2}$) & (AB mag)  & (10$^{41}$ erg s$^{-1}$) & (kms$^{-1}$) & (kpc) & (M$_\odot$ yr$^{-1}$) \\
        \hline
         \multicolumn{7}{c}{J2351+1600} \\
         \hline 
          54 & 23:51:51.6 & 16:00:56 & 18.2$\pm$0.3 & 1.27$\pm$0.17 & $>$28.40 & 1.80$\pm$0.24 & -252$\pm$16 & 132.4$\pm$2.4 & $<$0.37 \\ 
          86 & 23:51:53.7 & 16:00:58 & 17.2$\pm$0.8 & 2.60$\pm$0.29 & $>$28.09 & 3.68$\pm$0.41 & +188$\pm$16 & 124.9$\pm$5.5 & $<$0.49 \\ 
         \hline
        \multicolumn{7}{c}{J0851+2332} \\
         \hline 
          83* & 08:51:43.6 & 23:32:12 & 3.3$\pm$0.3 & 9.98$\pm$0.62 & 25.34$\pm$0.14 & 11.94$\pm$0.74 & +262$\pm$17 & 24.9$\pm$2.1 & 5.55$\pm$0.72 \\ 
          95 & 08:51:42.0 & 23:32:26 & 29.6$\pm$0.5 & 2.77$\pm$0.33 & $>$27.68 & 3.32$\pm$0.39 & +270$\pm$17 & 221.0$\pm$4.0 & $<$0.64 \\ 
         \hline
        \multicolumn{7}{c}{J0255+0048} \\
         \hline 
          56* & 02:55:18.8 & 00:48:49 & 4.0$\pm$0.5 & 30.99$\pm$1.18 & 24.58$\pm$0.22 & 30.53$\pm$1.16 & +215$\pm$18 & 30.9$\pm$3.6 & 9.81$\pm$1.96 \\ 
          78 & 02:55:16.7 & 00:49:06 & 33.7$\pm$1.6 & 4.72$\pm$0.46 & $>$27.46 & 4.65$\pm$0.45 & +343$\pm$18 & 259.0$\pm$12.1 & $<$0.69 \\ 
          108 & 02:55:18.7 & 00:48:57 & 10.9$\pm$0.6 & 3.18$\pm$0.38 & $>$27.36 & 3.13$\pm$0.37 & +774$\pm$18 & 83.8$\pm$4.5 & $<$0.76 \\ 
         \hline
        \multicolumn{7}{c}{J1220+0921} \\
         \hline 
          45* & 12:20:21.8 & 09:21:17 & 19.6$\pm$0.2 & 14.62$\pm$0.63 & $>$26.77 & 14.97$\pm$0.64 & -279$\pm$18 & 150.1$\pm$1.9 & $<$1.34 \\ 
          85* & 12:20:23.1 & 09:21:10 & 36.2$\pm$0.2 & 27.45$\pm$0.92 & 25.87$\pm$0.20 & 28.12$\pm$0.94 & +370$\pm$18 & 277.1$\pm$1.9 & 3.08$\pm$0.58 \\ 
          91 & 12:20:22.1 & 09:21:40 & 11.7$\pm$0.3 & 3.03$\pm$0.30 & $>$27.57 & 3.10$\pm$0.31 & +120$\pm$18 & 89.3$\pm$2.0 & $<$0.64 \\ 
          98* & 12:20:20.3 & 09:21:52 & 22.9$\pm$0.3 & 13.39$\pm$0.62 & $>$26.81 & 13.72$\pm$0.63 & +229$\pm$18 & 175.0$\pm$2.1 & $<$1.29 \\ 
         \hline
        \multicolumn{7}{c}{J0818+2631} \\
         \hline 
          87 & 08:18:14.5 & 26:31:47 & 22.9$\pm$1.0 & 3.24$\pm$0.42 & $>$26.98 & 3.96$\pm$0.51 & +232$\pm$17 & 170.7$\pm$7.4 & $<$1.24 \\ 
          91 & 08:18:13.5 & 26:31:33 & 7.1$\pm$2.0 & 4.87$\pm$0.56 & $>$27.11 & 5.96$\pm$0.68 & +291$\pm$17 & 52.9$\pm$14.5 & $<$1.10 \\ 
          119 & 08:18:11.3 & 26:31:18 & 29.1$\pm$0.7 & 4.79$\pm$0.44 & $>$27.23 & 5.86$\pm$0.54 & +799$\pm$17 & 216.9$\pm$5.2 & $<$0.99 \\ 
         \hline
	\end{tabular}
	\caption{Full list of DLA associations. * High confidence detections. $^a$  Position and velocities estimated using the \lya\ centroids. $^b$ \lya\ integrated flux measured from integrating inside the 3D segmentation map. $^c$ Upper limits use the apertures defined from the \lya\ segmentation map, and specifically pixels where the map includes at least two wavelength channels. $^d$ Values derived using the SFR calibration from \citet{fumagalli2010}, which assumes a Chabrier IMF.}
	\label{tab:LAEs}
\end{table*}

\subsubsection{J0851+2332}
A \lya-bright LBG was detected at +260$\pm$20 km s$^{-1}$ from the DLA redshift with an impact parameter of 25$\pm$2 kpc, as summarised in Table \ref{tab:LAEs}. Fig. \ref{fig:lya_SB} shows the continuum and \lya\ detection of this object. With an $r$-band magnitude of $25.34\pm0.14$~mag this galaxy is forming stars at a rate of $5.6\pm0.7$ \msun yr$^{-1}$. This value is calculated using the conversion presented in \citet{Madau1998} which has been re-normalised for a Chabrier IMF \citep{Salim2007, fumagalli2010}, and it is not corrected for any intrinsic dust obscuration. 
DLA J0851+2332 is a somewhat high-metallicity DLA for its redshift, for which LBGs associations have been previously detected (e.g. \citealt{Moller02}). As the velocity offset is estimated using the \lya\ emission line, which is commonly redshifted from the true systematic velocity due to the radiative transfer of \lya, it is quite plausible that the LBG and DLA are very close in velocity space. 
At $z\simeq 3$, LAEs detected at f$_{\rm Ly\alpha}$ $>1.5\times10^{17}$ $\rm erg~s^{-1}~cm^{-2}$ have a virial mass of $10^{10.9^{+0.5}_{-0.9}}~\rm M_\odot$ \citep{Gawiser07}, at the redshift of DLA J0851+2332 the virial radius of such a halo is 29.4 kpc. The galaxy detected in proximity to DLA J0851+2332 is however fainter (by $\simeq1.5\times$) than the lower limit of this sample. With an impact parameter of 25$\pm$2 kpc, the DLA has a projected distance from the DLA of the order of the virial radius. While there will be a large scatter in the halo mass of a single LAE, this does suggest that the DLA may be directly related to a fainter, undetected galaxy or extended halo gas. This is motivated by predictions from simulations which indicate the median impact parameter of a DLA and the true host is around 0.1 virial radii \citep{RahmatiSchaye}.

\subsubsection{J0255+0048}

The host for DLA J0255+0048 was discussed at length in \cite{fumagalli2017}, its detection is summarised in Fig. \ref{fig:lya_SB}. The extended \lya\ structure spans $37\pm1$ kpc along its major axis and is dominated by two clumps. While most of the source has no broadband counterpart, a compact continuum source is embedded towards the edge of this structure. Although the MUSE spectrum of this continuum source is noisy, this source has spectrophotometry consistent with an LBG at $z_{LBG}\simeq z_{DLA}$ (see Appendix \ref{ap:LBGs}). Thus, given its location between the \lya\ structure and the DLA at the same redshift, it is quite likely that the LBG forms part of the same structure. \footnote{The SFR for the continuum source reported in Table \ref{tab:LAEs} is higher than reported in \cite{fumagalli2017}, as we measure $r$-band magnitudes from the MUSE data and not from the Keck LRIS imaging.}

As shown in \cite{fumagalli2017} the double peak of the \lya\ emission line for this object stems from a velocity offset between the two clumps, with a separation of $\Delta v=140\pm20$ km s$^{-1}$. This velocity difference is consistent with the velocity offset of the two components seen in the DLA absorption lines, such as \SiII\ shown in Fig. \ref{fig:metal_lines} ($\Delta v = 155\pm6$  km s$^{-1}$). It was argued that the morphology of this source and the correspondence between the two components in absorption and \lya\ emission may hint that this system is a merger, which has triggered starbursts in two galaxies embedded in the clumps. Alternatively the clumps could be part of some extended collapsing proto-disk, although the scale of this system is difficult to reconcile with this picture.  Cycle 25 HST WFC3/IR observations (PID: 15283, PI Mackenzie) will soon offer more details on the nature of this system.

\subsubsection{J1220+0921}

Three high-confidence LAEs were detected in the field of DLA J1220+0921, and all three lie within $\pm400$  km s$^{-1}$ of the DLA redshift. With impact parameters between 150.4 to 278.2 kpc, these associations are unlikely to be the galaxies that give rise to the DLA system, but they trace the large scale structure in which DLA J1220+0921 is embedded in. Additionally, a lower significance LAE is detected in this field, id91. This candidate is much closer to the line of sight than the high-confidence detections at 89$\pm$2 kpc, with a velocity offset from the absorber of +120$\pm$20 km s$^{-1}$, and may be more closely associated with the DLA. However, \cite{PerezRafols18b} indicates that that lower metallicity DLAs have a characteristic halo mass of $9 \times 10^{10}$ \msun, which would imply a virial radius of $\sim$ 30 kpc. For this reason it is unlikely that any of the detected LAEs are directly connected with the DLA as they are at much larger impact parameters. With a metallicity of $Z/Z_{\odot}=-2.33$, DLA J1220+0921 is the most metal-poor DLA in our sample, and our MUSE observations reveal for the first time associations with a truly metal-poor DLA at high redshift.

Galaxy id85 is the brightest detection both in \lya\ and the $r$ band. Fig. \ref{fig:lya_SB} shows the \lya\ halo of id85 extends far beyond the UV continuum. This object has an impact parameter of 278.2 kpc (36.4 arcsec), it is offset from the absorption system by +370$\pm20$ km s$^{-1}$, and is forming stars at a rate of $3.1 \pm 0.6$ \msun yr$^{-1}$ based on the UV luminosity. The rest-frame UV spectrum of this object is consistent with an LBG with $z_{LBG}\simeq z_{DLA}$=3.309 (see Fig. \ref{fig:LBG_spec} lower panel), the Lyman break is convincingly detected but the spectrum is too noisy to detect the interstellar lines. The broadband spectrum combined with the \lya\ halo of this object make its redshift unmistakable. 
The other two LAEs, instead, do not show continuum counterparts at the depth of these observations.

Fig. \ref{fig:map_19H7} shows the distribution of the detected LAEs across the MUSE field of view, spanning approximately 50~arcsec diagonally across the image. Three detections and the DLA, at the position of the quasar, are roughly joined by a line, suggestive that these objects trace a filament within which the DLA is embedded. 
A similar configuration was detected with MUSE for a very metal-poor (pristine) Lyman Limit System, with multiple LAEs detected in filament-like arrangement around the absorber \citep{fumagalliLLS}. With J1220+0921 the spread in velocity space is smaller by a factor of $\sim2.5$ and the \lya\ luminosities are higher. Narrowband studies of emission line galaxies around DLAs \citep{FynboBridgeI} have also identified a large over-density of galaxies in proximity to a metal-poor DLA ([M/H]$\simeq-1.7$). \cite{GroveBridgeII} reported 23 emission line galaxies within $\sim8$ comoving Mpc of the sightline with a velocity dispersion of only 470 km s$^{-1}$, much narrower than the profile of the narrowband filter. 
To gain more insight into the nature of this system and what type of structure we might be observing, we search for analogues in the \eagle\ simulations, specifically in the 25 Mpc box with DLAs identified using {\sc urchin}. Fig. \ref{fig:map_19H7} (right) shows a projected \HI\ column density map over the same field of view as the one probed by MUSE at $z=3.25$. The figure is centred on a DLA pixel, selected because of its likeness to J1220+0921. Specifically, we searched for DLAs with 3 LAEs within the area defined by the MUSE FoV, but none within the inner $15\times 15~\rm arcsec^2$. In this search, we require that LAEs have luminosity $\rm L_{Ly \alpha} \ge 10^{42}$ erg s$^{-1}$ sub-sampled to match the luminosity function as described in Sect. \ref{sec:EAGLE_LAEs}. In the {\sc eagle} 100 Mpc simulation 1.7\% of DLA pixels match this selection criteria for a single realisation of the LAE sub-sampling. Out of all the matches, the example shown in Fig. \ref{fig:map_19H7} is selected due to its morphological similarity in the distribution of LAEs around the DLA. In this case, the DLA arises from a small galaxy at close impact parameter, which is in turn embedded in a filamentary structure hosting additional \lya\ bright galaxies. A wider view of the selected region further reveals that this whole structure is just part of a filament extending beyond the scale probed by of MUSE-like observations. 

While analogues to this system in the \eagle\ simulation support the idea of a filamentary structure, 
we note however that this picture is complicated by the large velocity offset between id85 and id45 of 650$\pm$25 km s$^{-1}$, which may be too large to be explained with the associations embedded in a single filament. 
Therefore, we cannot exclude that galaxies are instead embedded in a proto-group or cluster type environment, but not yet bound to the same halo.  

\subsection{Continuum counterparts of the LAEs}

As Fig. \ref{fig:lya_SB} shows, three of the five LAEs detected have an $r$-band counterpart visible at the depth of our MUSE observations. We have extracted spectra for these objects, which we present in Fig. \ref{fig:LBG_spec}. 
With integral field spectroscopy, we can also examine the nature of the candidate DLA hosts identified in \cite{fumagalli2015} based on impact parameters in deep $u-$band images. None of these sources are confirmed to be near the redshifts of the DLAs with the MUSE data, with most of them being in fact low-redshift interlopers (see Appendix \ref{ap:field_discuss}). This highlights the power of searching for DLA host galaxies with \lya\, rather than relying on broadband detections, and emphasises once more the perils of relying on proximity to the quasar sightline as the way to identify candidate hosts of DLAs.

\section{Discussion} \label{sec:discussion}

Following the analysis of the observations in Sect. \ref{sec:data_reduct} and Sect. \ref{sec:GalSearch}, we have described the models used in this paper in Sect. \ref{sec:models}. Starting with the \eagle\ simulations, we have derived DLAs either by post-processing the simulation box with the {\sc urchin} radiative-transfer code (hereafter {\sc urchin} model), or by ``painting'' DLAs to halos from the simulations using a simple prescription that can be calibrated to match the large-scale bias of DLAs (hereafter painted models). We have also introduced two  models for LAEs, one simple model in EAGLE and one based on a semi-analytic prescription, both of which are calibrated to match the luminosity function of LAEs.

We now turn the discussion of our observations in the context of previous searches of DLA associations (Sect. \ref{sec:disc_obs}), moving next to the interpretation of our observations with models (Sect. \ref{sec:disc_mod}), and concluding with forecasts of how future searches will refine the determination of the typical properties of DLA hosts via small-scale clustering of Ly$\alpha$ emitters (Sect. \ref{sec:forecast}).

\begin{figure}
	\centering
\includegraphics[width=0.5\textwidth]{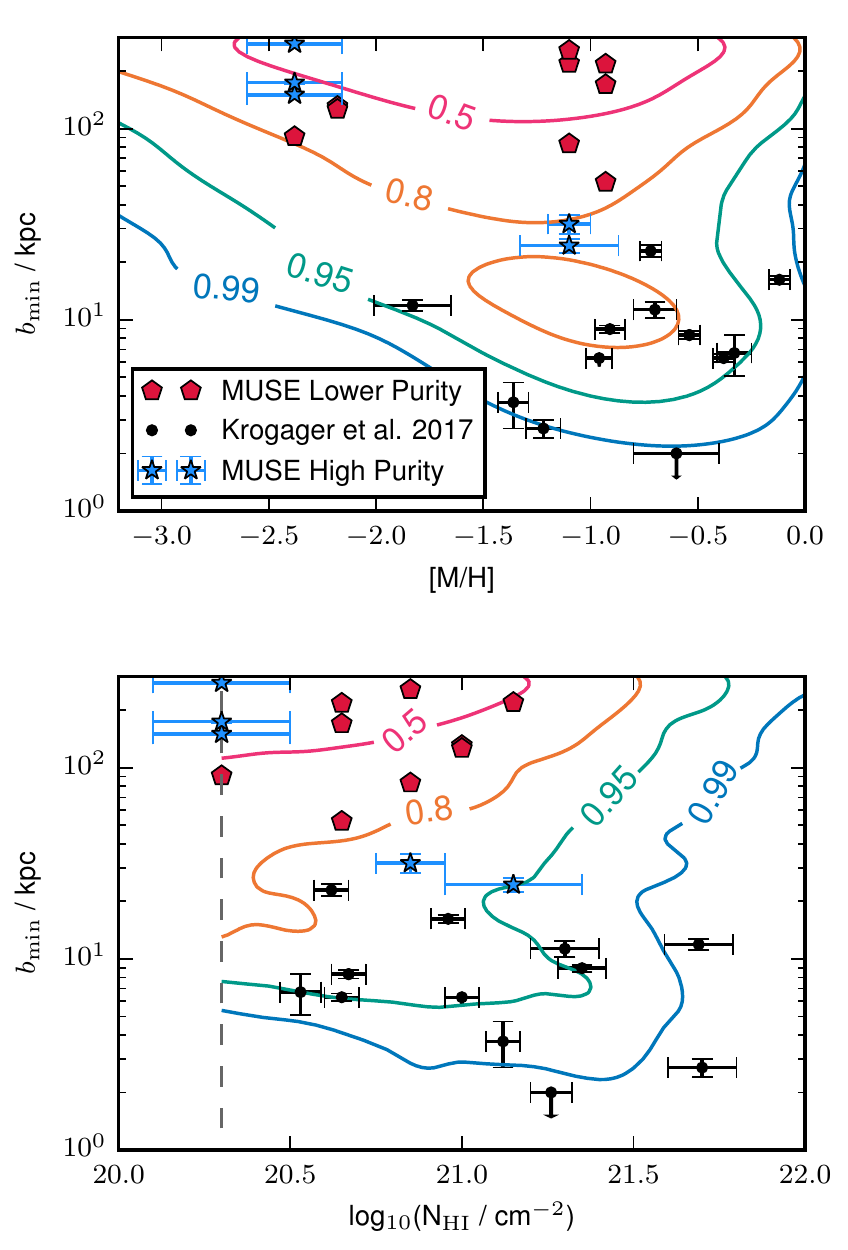}
	\caption{MUSE detected DLA associations in context with current samples of $z>1.5$ DLA host galaxies. The figure shows the galaxy impact parameter (in physical kpc) as a function of the DLA metallicity (top) and column density (bottom). The comparison sample (compiled in \citealp{Krogager17}, black points) and MUSE high confidence detections (blue stars) are plotted. The contours show predictions of the distribution of LAEs from the \eagle\ simulation using URCHIN DLAs, they are labelled by the enclosed fraction of total DLA-LAE pairs (within the plotting boundaries). The simulated contours apply for a MUSE like selection and do not simulate the varied selection of the literature sample. Both high and low purity detections are plotted. Note the simulated and observed MUSE comparison includes all LAE-DLA pairs and therefore the distributions are different to previous searches which have only considered the closest or brightest detection (e.g.\citealp{RahmatiSchaye}). In the lower panel the DLA column density threshold is marked (vertical dashed line).}
	\label{fig:prev_hosts}
	\vspace{-1em}
\end{figure}

\subsection{Detection rates and comparison with previous studies} \label{sec:disc_obs}

Previous searches for DLA host galaxies have revealed fewer than 20 spectroscopically confirmed DLA host galaxies at $z>2$. These have been identified over 25 years by a range of surveys with many different instruments including a highly-successful campaign with X-Shooter \citep[see a summary in][]{Krogager17}. Our MUSE survey, in under 15 hours of observations, has uncovered an unprecedented large-sample of associations with \lya\ emitters covering a larger field of view, detecting high-purity objects in three out of six targeted DLA fields, and more systems (although with lower confidence) in five of six fields we have observed. 
As discussed in Sect.~\ref{sec:GalSearch}, it is difficult to cleanly separate the lower confidence LAEs from contaminants, but based on the their clustering in velocity space it is very likely many of these are real DLA associations.

Taking only the high-purity detections as the lower limit, we establish a detection rate of at least $50\%$, which rises to 60\% if one excludes the field J0818+2631 which suffered from shallower and incomplete observations. While we have been extremely successful in discovering associated galaxies, we have only detected two plausible host galaxies. This detection rate is consistent with surveys that did not pre-select targets in metallicity. \citep{Moller02} This detection rate is considerably lower than what found in metallicity selected samples (e.g. {\citealt{Krogager17}, 64\%), but our observations have the advantage of enabling the study of the clustering properties of the full DLA population.

This work therefore confirms the competitiveness of \lya\ as a means to search associated galaxies in the DLA environment.
As noted, we are searching to larger impact parameters than some of the previous surveys, and have detected in some cases galaxies at sufficiently large impact parameters to make unlikely a direct connection between most of our detected LAEs and the gas measured in absorption. 
\lya\ also appears a powerful complementary technique to searches with ALMA at high metallicity \citep{neeleman2017,Neeleman2019}, as obscured systems where \lya\ would be absent due to scattering can be revealed instead via FIR dust continuum and [CII] emission. While published detections with ALMA are currently confined to highly star forming galaxies (5 and 110 \msun\ yr$^{-1}$), MUSE enables the detection of SFRs of the order of 1 $\rm M_{\odot}$ yr$^{-1}$ and, hence, is sensitive to the lower SFR galaxies with low dust extinction that, arguably, constitute the bulk of the DLA hosts. Our detections, with fluxes $\gtrsim 10^{-18}~\rm erg~s^{-1}~cm^{-2}$ and luminosities of  $\gtrsim 10^{41}~\rm erg~s^{-1}$, overlap with the population of faint \lya\ emitters detected in deep long-slit observations by \citet{Rauch}.
It is therefore quite plausible that our programme has in fact  finally detected within quasar fields the tip-of-the-iceberg of the faint population of LAEs where DLAs originate. As argued in \citet{Rauch}, these small proto-galactic clumps have individually only limited cross section, but are numerous enough to explain the abundance of DLAs, which in turn is consistent with the number density of faint LAEs \citep{Leclercq17,Wisotzki18}. 

As our search differs from many previous efforts in that we are not limited to small-impact parameters due to the large MUSE field of view, nor did we pre-select fields based on absorption properties, 
it is interesting to compare our detected associations with the existing sample of spectroscopically confirmed host galaxies. Fig. \ref{fig:prev_hosts} shows a comparison between the MUSE detected associations and the DLA host galaxies compiled in \cite{Krogager17}, plotted as impact parameter against the metallicity and column density of the DLA. 
In the case of the MUSE associations, all detections are shown including both the high-purity and lower confidence sample. Also shown are contours of the distribution of DLA-LAE pairs as simulated with \eagle\ using URCHIN DLAs. The contours enclose the fraction of the total simulated DLA-LAE pairs which lie within a defined region of parameter space. For example the red contour in the lower plot (and the plot boundaries and DLA threshold) encloses 50\% of all simulated DLA-LAE pairs with $1<b_{\rm min}<250$ kpc and $2 \times 10^{20} < \rm N_{HI} < 10^{22}$ cm$^{-2}$. The pair counts include DLAs with multiple LAEs within 250 kpc, not simply the closest match, and vice versa. For our MUSE sample we would expect 50\% of pairs (with $b_{min}<250$ kpc) to lie in the red contour, and 80\% in the orange contour. When the lower purity sample are included these fractions agree reasonably well.

The first observation is that indeed the MUSE detected associations are almost exclusively at larger impact parameters than the literature hosts, as it is also the case for the ALMA sample \citep{neeleman2017}. It is expected that $b_{\rm min}$ will typically be larger in the case of the MUSE observations than the literature sample, as we have plotted associated galaxies even in the cases where it is unlikely there is a direct link with the absorbing gas. However, that does not immediately explain why all our detections would be found at larger impact parameters
compared to previous searches, it may be due in part to including low \NHI\ and low metallicity systems \citep{fumagalli2015}. We argue instead that our search, although not revealing in many case the direct host galaxies which fall below the detection limit, provide a more representative view of the typical galaxy population around DLAs. Previous searches, either by virtue of the detection method or by the fact the only the closest galaxies may have been reported as DLA hosts in some instances, are likely to have yielded samples that include only the brightest and closest associations without capturing the full or even more typical distribution of the properties of galaxies near DLAs. Two distinct regions can be identified in the \eagle\ contours for both metallicity and column density, one population with $b_{min} <$ 20 kpc and a broader population for 
close associations which increases towards increasing $b_{min}$. The $b_{min} <$ 20 kpc are presumably host galaxies.  Indeed, only when we include all detected sources the top part of Fig. \ref{fig:prev_hosts} (the locus of associations) becomes highly populated.
The two detections from the MUSE results which lie closest to the locus of host galaxies are the high confidence detections in J0255+0048 and J0851+2332. 

From the top panel of Fig. \ref{fig:prev_hosts}, there is no obvious correlation between impact parameter and metallicity in detected sources. It is indeed believed that there is only a weak relation between the two properties, which is only apparent when controlling for other factors \citep{Christensen2014}. The lower panel shows instead a stronger anti-correlation between impact parameter and column density, as reported in \cite{Krogager17} and \cite{Rhodin2018} and the references therein. Taking only the high purity sample and neglecting J1220+0921 (as none of the detected galaxies would be included in the literature sample as hosts), the Pearson's correlation coefficient between log$_{10}$( \NHI) and log(b$_{min}$) is -0.66, which has a p value of 0.034. This significance is unchanged from \cite{Krogager17} by the addition of two host candidates presented in this paper. This is also consistent with the predictions from simulations \citep{RahmatiSchaye}.

\subsection{Constraints on the DLA host halo mass}  \label{sec:disc_mod}

As discussed in the previous sections, in most cases we have no evidence that the detected associations correspond to the DLA host galaxies, but rather the detected LAEs act as tracers of the environments within which the DLAs arise. Nevertheless, within a given cosmological model, we can still constrain the properties of the host galaxies (and in particular their typical halo mass) by comparing the small-scale clustering of LAEs with predictions as a function of halo mass. This approach is similar to the DLA/LBG cross-correlation analysis by \citet{Cooke2006}, albeit on smaller scales, and further offers additional insight into the high-bias reported by \cite{BOSS2018}

To this end, we first calculate radial density profiles of LAEs around DLA sightlines. These profiles are constructed by selecting a DLA and counting the mean number of LAEs inside a velocity window and given radius $\rm D_{LOS}$, as a function of $\rm D_{LOS}$. This is shown in Fig. \ref{fig:N_r}, where the MUSE observations are compared against both {\sc urchin} DLAs and the results of the painted halo scheme, applied to both the \eagle\ 100 Mpc simulation and the semi-analytic model. The painted model shown in this case has $\alpha$=0.75, using the parameters shown in Table \ref{tab:BOSS_param}.  We use the high confidence sample for this task as there are uncertainties in the purity and completeness of the lower SNR candidates. As the high-purity LAEs extend down only to a luminosity of $\approx 10^{42}$ erg s$^{-1}$, we apply this limit also to the modelled LAEs. We empirically estimate the uncertainty in the measured profile using jackknife resampling. To do this, we calculate the radial profiles (N$_{\rm gal}(<D_{LOS})$ or $\rho(<D_{LOS})/\rho$) for subsamples which omit the $i^{\rm th}$ sightline. From these subsamples and mean radial profile we estimate the variance on our observed radial profile using Eq. \ref{eq:jackknife},

\begin{equation}
\sigma_x^2 = \frac{n-1}{n} \sum^n_{i=1} (\Bar{x}-x_i)^2,
\label{eq:jackknife}
\end{equation}

where in $n$ is the number of subsamples (in this case the number of sightlines, six) and $x$ is the radial density profile. Had we used a more traditional estimator of the clustering the errors would be  larger, given that the profiles we are using are cumulative. In other words, jackknife resampling gives us an empirical estimate of the uncertainty on the radial profile. As we will show in Sect. \ref{sec:forecast} these jackknife errors may be lower but, they are comparable to the true uncertainty. As at D$_{LOS} < 24$ kpc there are no detections, we place an upper limit on the profile by calculating the maximum rate at which we would expect at least one detection over six DLAs with 66.7\%\ confidence assuming Poissonian statistics. 

Based on our, admittedly large, empirical uncertainties both the painted halos and {\sc urchin} DLAs are consistent with our observations, perhaps with a preference for the painted models, implying that models of LAEs clustered within dark matter halos are able to reproduce the observed radial profile. Here, we take the velocity window to be $\pm$1000 \kms, as this corresponds to the length of 25 Mpc at $z=3.4$ (the mean DLA redshift) which matches the size of the \eagle\ 25 Mpc simulation.

\begin{figure}
	\centering
	\includegraphics[width=0.5\textwidth]{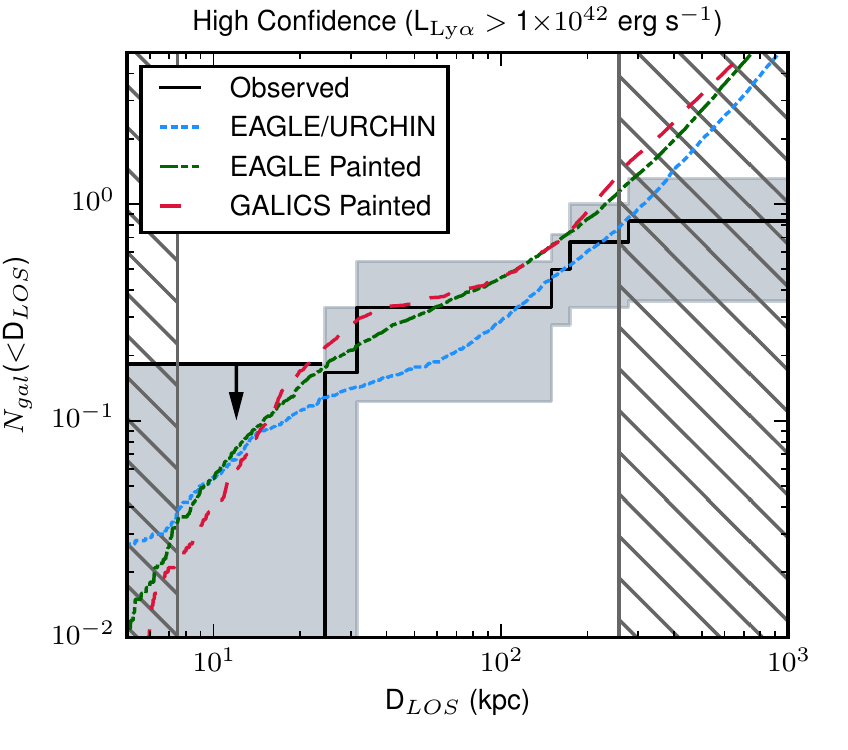}
	\caption{The cumulative distribution of LAEs around DLA sightlines for our MUSE observations (high confidence only, black) and two simulated datasets. N$_{gal}$ is the average number of bright LAEs within a radius (D$_{LOS}$) and a velocity window of 1000 \kms from a DLA.  The DLAs used in calculating the simulated profiles come from \eagle\ with {\sc urchin} post-processing  (blue, dotted) and the painted halos applied to both the \eagle\ 100 Mpc (green, dot-dahsed) and \GALICS\ (red dashed) data. The 1$\sigma$ error in the profile from the observations is estimated via jackknife resampling (grey shaded), and, at small scales where there are no detections, the 66.7\%\ upper limit is shown (black arrow). The grey hatched regions indicate the limits of scales probed in the MUSE observations, between 1 arcsec (of the order of the seeing) and 0.5 arcmin (the field of view) converted into a proper distance at the typical DLA redshift.}
	\label{fig:N_r}
	\vspace{-1em}
\end{figure}

Next, to remove the geometric effect of greater volume as $\rm D_{LOS}$ increases, we convert the number of LAEs within $\rm D_{LOS}$ to a density measured in a cylindrical aperture. Additionally we divide these density profiles by the mean number density of LAEs, to convert the density profiles in terms of over-density of LAEs with respect to the mean density. In the case of the \eagle\ simulations the mean number density is simply calculated from the model, while for our MUSE observations and the \GALICS\ mock catalog, the mean density is taken from integrating the \cite{Drake2017} Schechter luminosity function down to \lya\ luminosities of $\rm 10^{42}$ erg s$^{-1}$. 

\begin{figure*}
	\centering
\includegraphics[width=\textwidth]{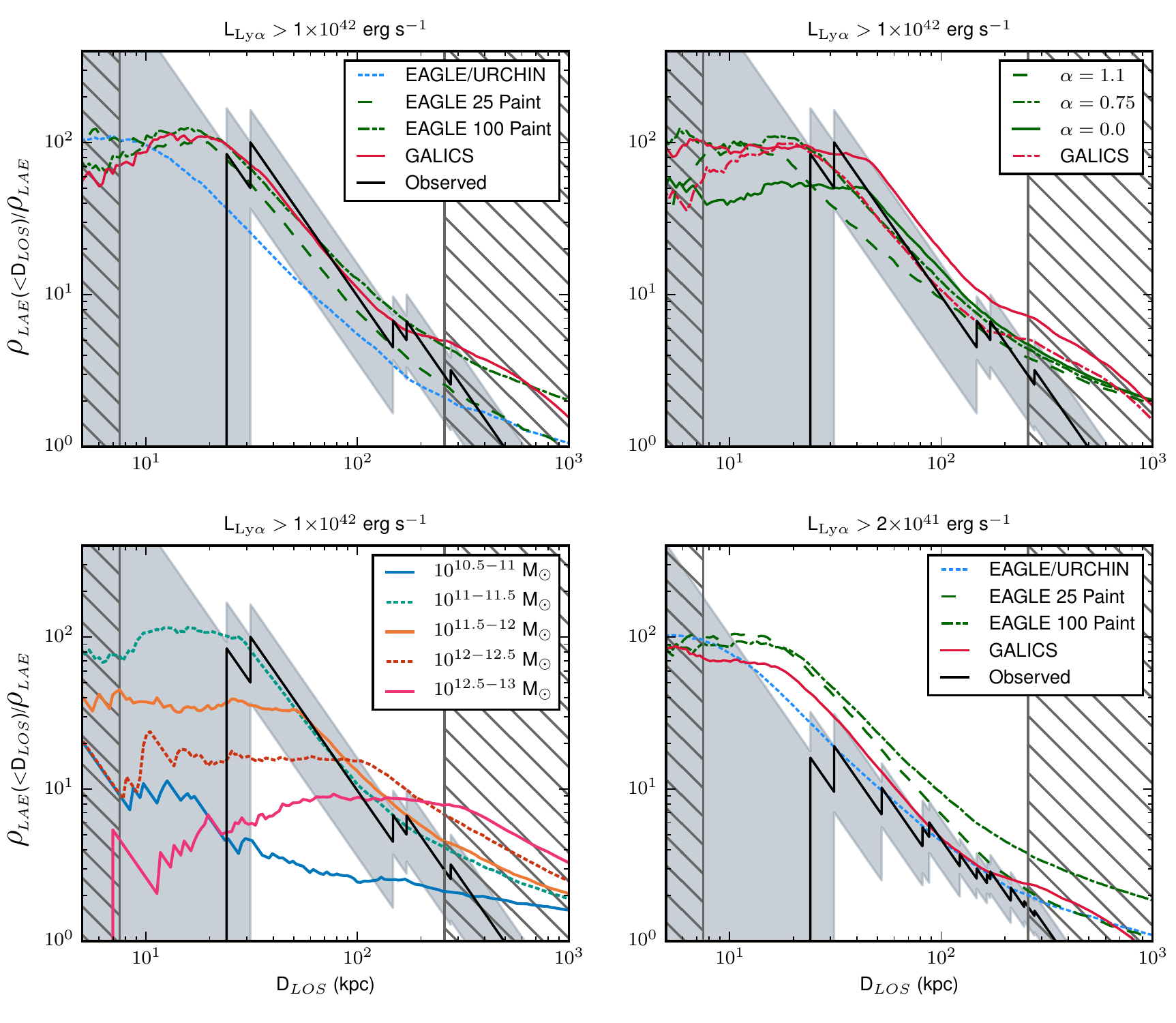}
	\caption{The radial over-density profile of LAEs around DLAs, computed as the density of LAEs with respect to the mean density, measured as a function of the circular aperture $\rm D_{LOS}$ from a DLA, simulated and observed. Vertical dashed lines indicate 1 arcsec and 0.5 arcmin, respectively, converted in physical distances at the mean redshift of the MUSE DLA sample. 
    	\textbf{Top left:} Shown are our MUSE results (black, solid) in comparison to the DLAs defined by {\sc urchin} post-processing of the \eagle\ 25 Mpc box and the painting model that matches the \protect\cite{BOSS2018} results. The painted DLA are shown for both the \GALICS\ mock catalog (red, solid) and the \eagle\ 25 and 100 Mpc simulations (green, dot-dashed and dashed lines respectively). The shaded region shows the 1 $\sigma$ error in the observed radial profile estimated though resampling, with a calculated upper limit at small scales.
		\textbf{Top right:} The MUSE observations are shown against results from the \eagle\ 100 Mpc box with painted DLAs shown with 3 different model parameters ($\alpha=1.1, 0.75$ and 0.0). The profile from the \GALICS\ lightcone are also shown (red), but only for $\alpha=0.75$ and $0.0$.
		\textbf{Lower left:}  Halos from the \eagle\ 100 Mpc box are selected by halo mass ($M_{\rm h}$) and painted with a uniform cross-section. The selected ranges are shown in the legend. The 10$^{11}$-10$^{11.5}$ \msun that best matches the observed profile is similar to the characteristic halo mass estimated from DLA clustering, however in reality DLAs will have a distribution of halo masses.
		\textbf{Lower right:}  The same plot as the top right but now using all detected LAEs down to a luminosity of $2\times10^{41}$ erg s$^{-1}$, i.e. 12 of 13 candidates and confirmed detections. The completeness estimated for the MUSE data has been applied to the simulations rather than as a correction to the data. At this lower luminosity, the observations now agree more closely with the \eagle/{\sc urchin} prediction,  but this result is more uncertain due to difficulties in estimating the LAE completeness.}
	\label{fig:profile_sim}
	\vspace{-1em}
\end{figure*}

\begin{figure}
	\centering
	\includegraphics[width=0.5\textwidth]{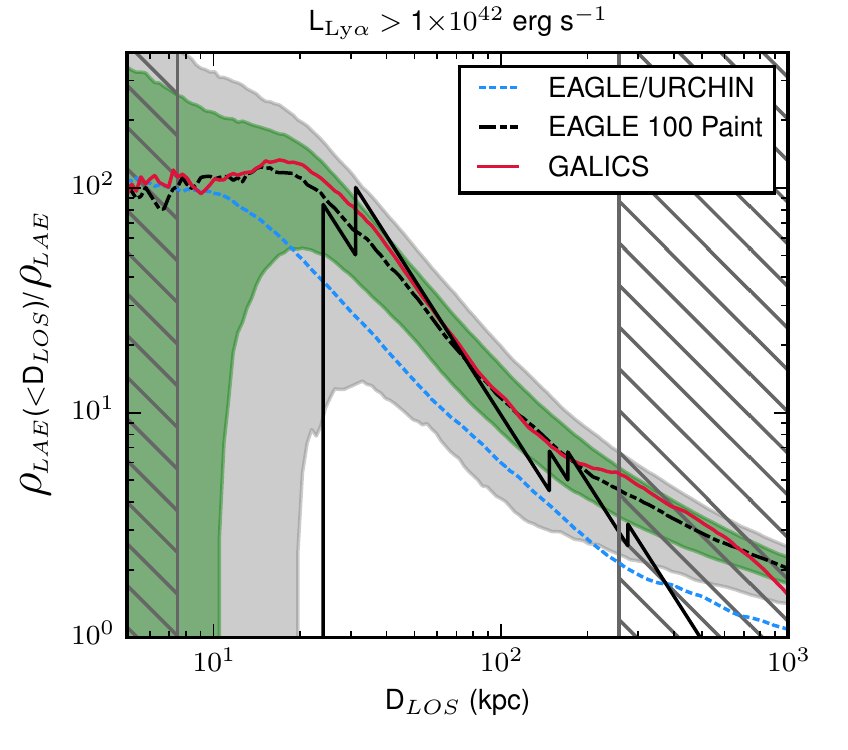}
	\caption{The simulated scatter of sets of 6 DLA with MUSE-like observations from the EAGLE 100 Mpc painted simulation (grey shaded). The error that would be obtained with a hypothetical 24 DLA MUSE survey is also displayed (green shaded).}
	\label{fig:forecast}
	\vspace{-1em}
\end{figure}

Fig. \ref{fig:profile_sim} (top left) shows the relative density profile for the MUSE DLA observations in comparison to both model DLAs identified with {\sc urchin} and from the painted a halo model. The halo model used here is $\alpha$=0.75, which is the same for all three painted simulations, the \GALICS\ mock catalog and the \eagle\ 25 and 100 Mpc boxes. We use this intermediate model as it corresponds to halo masses which are suitably converged in all three simulations. The shape of the relative density profile probes several properties of the DLA population. The initial plateau of the curves at small radii describes the extent of individual DLAs, while the tail encodes small-scale clustering information. 

Inspecting this panel, the most notable difference is between the density profile from the {\sc urchin} DLAs and those painted  with the halo prescription: the density at larger scales is much less enhanced for {\sc urchin} DLAs. This is because, within \eagle, DLAs populate also small halos due to the lack of a low-mass cut-off, and are thus less clustered on average. We also observe that the painted model and results from the {\sc urchin} calculation in the \eagle\ 25 Mpc  converge at large scales, as expected by construction given that they are the same box and they should converge to the same mean density. 
We note that some of the suppression of {\sc urchin} results with respect to the 100 Mpc simulation and \GALICS\ may be due to limited volume, as the 25 Mpc painted DLAs lie below the two larger painted simulations.

We investigate next whether the discrepancy with the {\sc urchin} DLAs is significant.
The good agreement between the 100 Mpc \eagle\ box and the \GALICS\ lightcone with painted DLAs strengthens these results, showing that the assumptions made about LAEs in \eagle\ are sufficiently robust for this comparison. From Fig. \ref{fig:profile_sim}, the most substantial conclusion is that the MUSE observations appear to be in good agreement with the halo prescription of \cite{FontRibera12}. In this paper we do not attempt to investigate the metallicity dependence of DLA clustering reported by \cite{PerezRafols18b} as we have few DLA fields, but future studies with larger samples may be able to probe trends with metallicity.

In Fig. \ref{fig:profile_sim} (top right), the relative density profiles for the three sets of parameters for the halo prescription (Eq. \ref{eq:halomod}, Table \ref{tab:BOSS_param}) are shown when applied to the EAGLE 25 Mpc box (green). For $\alpha=0.75$ and $0.0$, the profile from the \GALICS\ mock is also shown (red). \GALICS\ suffers from limited mass resolution below halo masses of $\approx 2\times 10^{9}$ \msun\ and limited volume at high masses ($> 2 \times 10^{12}$ \msun\ ) so we only apply two lower $\alpha$ models which have a high enough cut-off mass. As described above, in each case, the absolute normalisation ($\Sigma_0$) is fixed to reproduce the DLA number density, $\ell(X)$, derived with {\sc urchin} post-processing in the 25 Mpc \eagle\ simulation. At large scales ($>100$ kpc), the three models are indistinguishable, as it is expected given that all three sets of parameters are chosen to match the observed bias of DLAs measured at large scales. The largest difference between the different models is at scales of $\simeq20$ kpc, which is due to the changing distribution of DLA sizes affecting where the flat part of the curve ends. For the most extreme model ($\alpha=0$), the DLA cross-section is fixed for all halos above $\rm M_{min}$, and in this case the sharp break in the profile is set by this fixed DLA radius. 

The relative density profile is largely insensitive to these different painting models, despite the minimum mass raising from $3\times10^9$ to $2.2\times10^{11}$ \msun.  This is an indication that the relative density profile in the halo prescription is mainly dominated by the effect of the massive halos ($\rm M_{h}>$10$^{11}$ \msun), as shown more explicitly in Fig. \ref{fig:profile_sim} (lower left). In this panel, the contribution of different halos has been split into bins of halo mass, where in each bin the halos are painted with a constant cross-section (i.e. $\alpha$=0). As before, the constant $\Sigma_0$ is calibrated in the 100 Mpc simulation to match the $\ell(X)$ predicted from the {\sc urchin} calculation. In the lowest mass bin, the halos simply do not contain enough LAEs or cluster with them strongly. In the 10$^{11}<M_{h}<$10$^{11.5}$ \msun\ bin the profile reproduces the MUSE data well. Above 10$^{12}$ \msun\ the profile peak declines, because the DLA cross-sections have become so large that some chosen DLAs can be very far from the halo center. Fig. \ref{fig:profile_sim} shows that, in this painting scheme, the characteristic halo mass preferred by our observations is $10^{11}\lesssim M_{h} \lesssim 10^{12}$ \msun\, which interestingly agrees with the halo mass yielding an equivalent bias to that of inferred from DLAs ($\approx 10^{11.78}$ \msun\ , \citealt{BOSS2018}). This single mass model is clearly a simplification, and we can refine this estimate by returning to  Eq. \ref{eq:halomod} of the model rather than assuming a constant dependence with mass. Using the 100 Mpc \eagle\ simulation, we compute the mean halo mass of a DLA in this scheme, finding that, for the values of $\alpha$ we tested, this characteristic halo mass falls in the same range as before, ranging from $10^{11.27}$ ($\alpha=1.1$) to $10^{11.78}$ \msun\ ($\alpha=0$). Taken together, the convergence of these estimates on the characteristic halo mass corroborates the finding of a high bias inferred from the large-scale clustering measurement.

In the above comparisons, we have utilised only the high-purity LAE sample, as the fainter detections suffer from limited completeness and may lack purity. To test the robustness of our results, we now lower the threshold to include LAEs with L$_{\rm Ly\alpha} \geq 2\times10^{41}$ erg s$^{-1}$.  
As discussed in Appendix \ref{ap:robust}, there is limited completeness to this luminosity threshold, therefore we must apply a correction prior to comparison with simulated profiles. To do so, we have chosen to apply the estimated incompleteness function (derived from the mean completeness measured over all six fields using realistic mock sources) to the simulations, as this procedure is less noisy than weighting the data by completeness. Fig. \ref{fig:profile_sim} (lower right) shows the observations compared with the modelled results, demonstrating that at this lower luminosity, the higher number of objects has improved our constraints on the density profile. It can be seen that when considering these fainter candidates, the data now appears to match the {\sc urchin} and \GALICS\ profiles more closely, as opposed to the painted \eagle\ and \GALICS\ preferred by the higher luminosity data. Indeed, the data now weakly reject the painted model applied to the \eagle\ 100 Mpc simulation. However, one should remember that the data are highly correlated due to the profiles being cumulative, and so the apparent significance is exaggerated. 

This discrepancy between low and high luminosities could be understood if the \lya\ luminosity function differs around DLAs with respect to the field. As the disagreement extends to 200 kpc, this appears to be unlikely to be due DLA host galaxies alone, as this scale is far beyond the virial radii of typical halos at this redshift. However luminosity dependence in the clustering of LAEs has been observed \citep{Ouchi03, Khostovan18}, this would mean that the luminosity function would be differ between different large-scale environments. The more robust modelling of \lya\ in \GALICS\ compared to \eagle\ may capture this effect better and could explain why \GALICS\ matches both the higher and lower luminosity threshold results simultaneously. Another hypothesis is that the LAE completeness is lower than estimated and hence, the data are biased low with respect to the models. Indeed, as detailed in Appendix \ref{ap:robust}, our estimate of the completeness does not include the additional manual vetoing stage, which cannot be trivially reproduced in large samples. It is therefore likely the completeness measured is overestimated. This uncertainty justifies our previous choice to focus our analysis on the high confidence sample. We do emphasise, however, that our sample is still rather small, and therefore any current inference is likely to be subject to sample variance (see next section).

Fig. \ref{fig:profile_sim} also shows a divergence in model predictions with luminosity. At high luminosity (L$_{\rm Ly\alpha}> 10^{42}$ erg s$^{-1}$)  both the painted \eagle\ and \GALICS\ profiles are in very close agreement for the same model parameters. However, at the lower luminosity, the predictions start diverging despite being derived with the same DLA prescription. Future datasets will require more precise  modelling of LAE properties, and - conversely - deeper observations will be able to discriminate more among different models. 

If some significant fraction of the DLA population are indeed hosted in massive halos, one can ask why bright LBGs are not typically detected in past photometric searches for DLA host galaxies. Using the \eagle\ 100 Mpc simulation with the painting method described above, we calculate the fraction of the total DLA cross-section which is hosted by halos which also contain a UV bright galaxy ($\rm M_{FUV}<-20$). The FUV magnitudes are taken from the} \eagle\ database and are calculated based on the GALEX FUV filter in the rest-frame, including dust attenuation \citep{Camps2018}. Using the scheme parameters defined in Table \ref{tab:BOSS_param}, we find fractions of 0.03, 0.11 and 0.22 for $\alpha=$1.1, 0.75 and 0.0 respectively. This is only including LBGs in the same halo, i.e. the one halo term not accounting for LBGs which may be clustered on larger scales. We conclude that the low rate of LBG detections around DLAs is consistent with the models motivated by clustering measurements.

As discussed in Sect. \ref{sec:EAGLE_LAEs}, the clustering amplitude of our simulated LAEs is somewhat higher, both in \eagle\ and \GALICS, once compared  to observations. To assess the impact this systematic error has on our result we repeated the exercise shown in Fig. \ref{fig:profile_sim} but substituting the LAEs with tracers that have a lower clustering amplitude, in line with observations. We select halos with $\rm M_{halo}>10^{10.6}$ \msun, as motivated by \cite{Gawiser07} who noted this selection matched their observed $\rm r_0$. This experiment lowers the simulated $\rm \rho(<D_{LOS})/\rho$ profiles. For the predictions based on DLAs identified with {\sc urchin} this lower bias test decreases the profile by $\approx 30$\% at a scale of 100 kpc. There is some dependence on this factor with radius, but the suppression is relatively constant from 10 to 100 kpc and then decreases gradually at larger radii. In the case of the the painted \eagle\ results the decrease in the relative density profiles is $\approx 50$\% again at 100 kpc. Although a factor of 2 is substantial, it is currently smaller than our statistical uncertainties, and we therefore believe this ambiguity does not impact our conclusion that both {\sc urchin} and our painted scheme are consistent with the data. Extending this test to the models which use only a single halo mass (see Fig. \ref{fig:profile_sim}, lower left), we observe that the 10$^{11}<M_{h}<$10$^{11.5}$ \msun\ bin still matches our observations best, however the 10$^{10.5}<M_{h}<$10$^{11}$ \msun\ bin profile now becomes consistent with the data. While these uncertainties in the properties of LAEs prevent us from currently making strong statements about the halo masses of DLAs based purely on our observations, such low halo masses are not consistent with the observed large-scale clustering of DLAs.

\subsection{Forecasts for future searches} \label{sec:forecast}

In Fig. \ref{fig:forecast} we show the 1$\sigma$ uncertainty one would obtain with a sample of 24 DLA fields studied with MUSE (green shaded), showing that, with future observations, it may be possible to start discriminating between  models. Datasets of this size are achievable in the near term with MUSE. Interestingly, if the underlying radial density profile were as low as predicted by the \eagle\ simulation with {\sc urchin} post-processing, the BOSS halo painting prescriptions could be ruled out. Some degeneracy will, however, exist between different DLA models and the clustering of the LAEs used in this cross-correlation, but it should be possible to break this degeneracy with surveys of LAEs that measure the large-scale clustering properties (e.g. MUSE-Wide and HETDEX at higher luminosities; \citealt{MUSEWide}).

In Fig. \ref{fig:forecast} we also test for the procedure adopted to estimate errors in the relative-density profiles.
In comparing our results to those of simulations, we have quantified the expected uncertainty in the radial density profile using resampling. With a only 6 DLAs, this may still underestimate the errors in the radial density profile. To quantify more robustly the uncertainty arising from sample variance, we drawn sets of 6 DLAs from the \eagle\ 100 Mpc box and use them to estimate the radial density profile. Iterating over 500 realizations of 6 DLAs each, we derive the scatter expected between sets of 6 DLAs. 
To quantify the uncertainty we estimate the standard deviation over all realizations. For this task, we use the \eagle\ 100 Mpc box using DLAs painted onto halos with $\alpha=0.75$. We prefer the larger simulation, as it provides many more independent realisations than the 25 Mpc box. Fig. \ref{fig:forecast} shows the 1$\sigma$ uncertainty (grey shaded) estimated from this Monte Carlo simulation for a survey of 6 DLA fields with MUSE-like observations. The simulated uncertainty is indeed larger than the errors estimated through resampling, particularly at $b_{min}<10$~kpc where there are only two detections. However, at larger radii the two estimates of uncertainties appear to be comparable. Based on this, we conclude that currently our results are compatible with both models described in this paper.

\section{Summary and Conclusions}\label{sec:conclusion}

We report the results from the first MUSE spectroscopic survey of high-redshift ($z\gtrsim 3$) DLAs in six quasar fields. We have searched for galaxies associated with six $3.2<z<3.8$ DLAs with deep rest-frame UV integral field spectroscopy, finding five high-purity \lya\ emitting galaxies in close proximity to three DLAs, plus 9 additional lower-significance objects in five out of six fields. Two of these galaxies are close enough to the DLAs to plausibly be host galaxies, directly connected with the material observed in absorption. Together with a very high detection rate of LAEs (up to $\approx 80\%$) in a sample not pre-selected by absorption properties, we report the detection of several galaxies tracing the dense environment of a very metal-poor DLA at high redshift. This DLA, in the field J1220+0921, appears to be embedded in a filamentary structure, as traced by three bright LAEs stretching across the full MUSE field of view.

We have also compared our results to the predictions of recent hydrodynamical simulations and a simple prescription in which DLAs are painted inside dark matter halos, showing that our detections of multiple associations support a picture in which DLAs arise from neutral gas in proximity to several galaxies clustered within dark matter halos. We have also explored the small-scale clustering of DLAs, by comparing the density of LAEs detected around DLAs to cosmological models in which DLAs and galaxies populate dark matter halos. The results of our MUSE survey are consistent with the predictions from both the \eagle\ simulation with radiative transfer post-processing, and a simpler scheme which paints DLA cross-sections onto dark matter halos. 
 More quantitatively, our comparison with simulations shows that a simple halo prescription tuned to reproduce the large-scale clustering of DLAs also explain the number of LAEs around DLAs on small-scales in our observations. Based on this model, we conclude that at least some DLA hosts have a characteristic halo mass of  $10^{11} \lesssim M_{\rm h} \lesssim 10^{12}$ \msun.

Considering individual detections,  in the field J0255+0048, MUSE has unveiled an extended 37 kpc \lya\ structure, potentially tracing a major merger perhaps hosted in a massive halo. In the field of DLA J1220+0921 we have discovered a rich environment, with several LAEs surrounding the DLA. Moreover, in half of our sample, we identified faint LBGs in proximity to the DLAs (i.e. within the MUSE field and 1000 km s$^{-1}$, see appendix \ref{ap:LBGs}). Jointly, these lines of evidence point to somewhat massive halos as hosts to a significant fraction of the DLA population. Furthermore, these new observations clearly show the need for more advanced models that incorporate clustering beyond the more traditional models treating individual galaxy/DLA pairs. 
We finally showed how MUSE observations in a large sample of 24 DLAs will provide sufficient statistical power to discriminate among different models of connection between DLAs and galaxies, and refine the mass estimates of halos hosting DLAs. Given the efficiency of MUSE, it should be possible to assemble such a sample in the near future.

\section*{Acknowledgements}

We would like to thank S. McAlpine and P. Norberg for helpful discussions, and R.A. Jorgenson for providing the spectrum of J1220+0921. RM, MF and TT acknowledge support by the Science and Technology Facilities Council [grant number  ST/P000541/1]. This project has received funding from the European Research Council (ERC) under the European Union's Horizon 2020 research and innovation programme [grant agreement No 757535]. TT additionally acknowledges the Interuniversity  Attraction Poles Programme initiated by the Belgian Science Policy Office [AP P7/08 CHARM]. TG is grateful to the LABEX Lyon Institute of Origins [ANR-10-LABX-0066] of the Universit\'e de Lyon for its financial support within the programme `Investissements d'Avenir' [ANR-11-IDEX-0007] of the French government operated by the National Research Agency (ANR). TG additionally acknowledges support from the European Research Council under grant agreement ERC-stg-757258 (TRIPLE). JXP acknowledges support from the National Science Foundation (NSF) grants AST-1010004 and AST-1412981. This work is based on observations collected at the European Organisation for Astronomical Research  in the Southern Hemisphere under ESO programme ID 095.A-0051 and 096.A-0022. Some of ancillary data presented herein were obtained at the W.M. Keck Observatory and Magellan Telescopes at Las Campanas Observatory. This research made use of Astropy, a community-developed core Python package for Astronomy \citep{astropy}. For access to the codes and advanced data products used in this work, please contact the authors or visit \url{http://www.michelefumagalli.com/codes.html}. {\sc topcat}, a graphical tool for manipulating tabular data, was also utilised in this analysis \citep{TOPCAT}. 
Raw data are publicly available via the ESO Science Archive Facility. 







\bibliographystyle{mnras}
\bibliography{bibliography.bib}






\appendix
\section{Additional tests on the robustness of the LAE sample}\label{ap:robust}

\begin{figure}
	\includegraphics[width=0.5\textwidth]{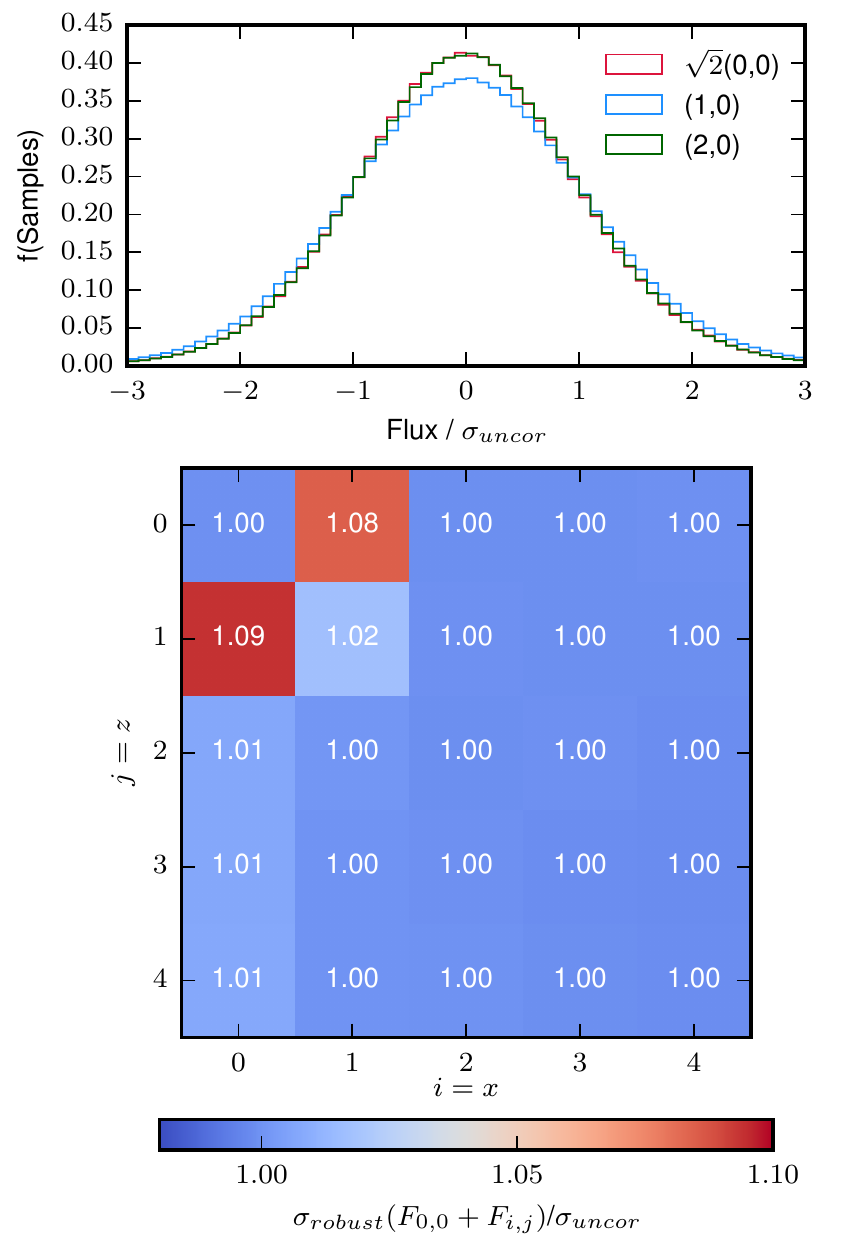}
	\caption{An example of the correlated noise in field J2351+1600 at the wavelengths corresponding to that of the search window. {\it Top:} The distribution of fluxes in pairs of neighbouring pixels normalized by the expected uncorrelated noise, showing the low level of correlated noise in the datacube. The normalized flux distribution for pixels that are immediate neighbours in the $x$ direction are shown in blue (1,0), as are the pixels separated by an additional 1 pixel in green (2,0). The expectation given uncorrelated noise, i.e. the flux distribution of single pixels multipled by $\sqrt{2}$, is also shown in red ($\sqrt{2}$(0,0)). {\it Lower:} The correlation matrix estimated from the datacube. The matrix shows the robust standard deviation obtained from the flux distribution from adding the cube to itself shifted in the x dimension by $i$ and the z direction by $j$. Each cell is coloured by the $\sigma$ and labelled with the value.}
	\label{fig:corr_noise}
\end{figure}

The significance of the detected LAEs is calculated from the rescaled variance of the pixels that enter the segmentation map. A possible concern is that, due to the the fact that detector pixels are resampled in the final datacubes, we may be overestimating the SNR by neglecting correlation in the noise between neighbouring pixels, as commonly noted in imaging \citep{fumagalli2015}. 
We test for this possibility in Fig. \ref{fig:corr_noise}, where we compare the flux distribution in the first and second neighbouring pixels to the theoretical value assuming uncorrelated noise. We also present the correlation matrix estimated from the datacube, computed from the standard deviation of pixels each added to their neighbouring pixels. If the noise is uncorrelated the expected value is $\sqrt{2}$ times the standard deviation of all pixels individually.  

Both tests show that following the resampling procedure some degree of correlation is introduced in the data, but this effect is limited to the nearest neighbours and thus it is unlikely to significantly alter our estimate of the SNR, which is believed to be accurate at the pixel following the rescaling described in the Sect. \ref{sec:LAEsearch}. 
This result is in line with the analysis of MUSE data presented by \citet{BaconUDF}. We note, however, that extended sources may have their SNR slightly overestimated due to the fact that we do not propagate covariance in our error estimate. 


\begin{figure}
	\includegraphics[width=0.5\textwidth]{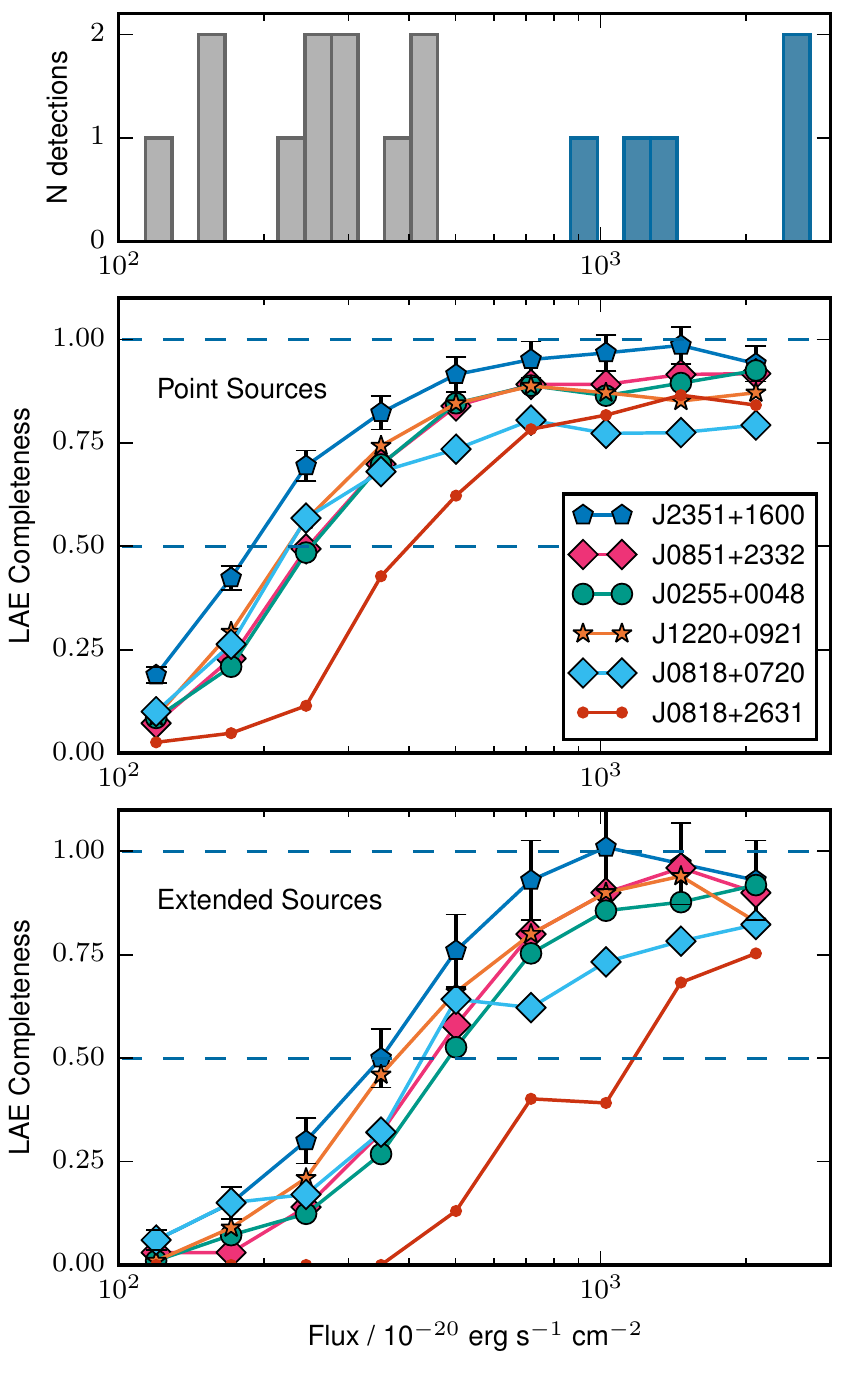}
	\caption{LAE completeness as a function of flux. Errors are shown for field J2351+1600. {\bf Middle:} the LAE completeness for different fields (black) shown against the \lya\ flux, based on injecting point sources. 50 and 100 \% completeness are indicated with blue dashed lines. In the case of field J2351+1600, the uncertainty in the completeness is also shown. {\bf Bottom:} The same as the middle panel but now for the realistic extended mock sources. {\bf Top:} The flux distribution of the high-confidence LAEs (blue) and lower significance candidates (grey)}
	\label{fig:completeness}
\end{figure}

We have also tested the completeness of our search to emission line objects. To this end, we inject mock emission lines in 
the detection datacubes, assuming a Gaussian profile both in the spectral (with 2 channels full-width at half-maximum) and spatial  directions (full-width at half-maximum of 3 pixels). This simulates unresolved point sources convolved with the seeing. We then run the {\sc CubExtractor} with the same settings adopted in our main search, and count the number of recovered mock lines with $\rm SNR \ge 7$ and $N_{voxels}\ge 40$, spanning more than 3 wavelength channels. Ten mock lines are injected at a time to limit crowding effects and possible blends, and the process is repeated 50 times for each cube and in different flux bins.  Additionally, we studied the effect of LAEs having extended profiles on the completeness. Following \cite{Drake2017}, we injected sources drawn from \cite{Leclercq17} which are rescaled in flux. This provides a more realistic estimate of the completeness, by using source parameters from a MUSE LAE sample. The sample presented in \cite{Leclercq17} had their Ly$\alpha$ surface brightness profiles fitted with two exponential profiles, representing an extended halo and a component which follows the UV continuum of the galaxies. The continuum-like components are typically unresolved by MUSE, so they are injected as Gaussian profiles with a FWHM matching the seeing. The flux ratio between components and the spectral FWHM are applied to the mock sources. We repeated this for ten iterations with ten mock sources per run. We believe this is a more accurate estimate of the completeness.

The estimated completeness as a function of line flux is shown in Fig. \ref{fig:completeness} for the 6 fields. This test shows that we are  $\simeq50\%$ complete for $\rm L_{Ly \alpha }=10^{41.5}$ erg s$^{-1}$ for point sources and $10^{41.8}$ erg s$^{-1}$ for realistic sources. This analysis, however, does not include the rejection of sources associated with the visual inspection, and is therefore a slightly optimistic estimate. From Fig. \ref{fig:completeness}, however, we conclude that the high-confidence LAEs (i.e. $\rm SNR=12$) do not suffer from incompleteness and form a highly-complete sample. The lower SNR candidates fall instead at lower completeness, although the exact value is uncertain due to the unknown effects introduced by the visual inspection.

\section{Notes on the fields with lower significance detections}\label{ap:field_discuss}

In the following sections, we individually discuss the detected LAEs in each DLA field where lower SNR objects have been found. The results from fields J0851+2332, J0255+0048 and J1220+0921 (containing high-confidence associations) are described in Sec. \ref{sec:results}. We also comment on candidate hosts identified in some fields from deep broadband imaging \citep{fumagalli2014} which we extract spectra for.

\subsubsection{J2351+1600}

Three low confidence candidates are detected in the field J2351+1600. These objects are offset in velocity from the DLA by 187.9, -32.3 and -252.0 km s$^{-1}$ in order from highest SNR, to the lowest. These candidates have impact parameters ranging from 130 to 166 kpc, which is likely too large to be directly associated with the gas giving rise to the damped absorber. This DLA was noted in \cite{fumagalli2015} as one of a 6 DLAs with candidate broadband counterparts, however MUSE observations of this object revealed it to be a $z=0.3886$ star-forming galaxy. 

\subsubsection{J0255+0048}
  
\cite{fumagalli2015} noted that in this field there is apparent emission at the position of the DLA, as measured in B band photometry. However, contamination from leakage of the quasar though the ``blocking'' Lyman Limit System could not be excluded. We do not detect \lya\ emission at the position of the quasar, however if there were a continuum object without \lya\ emission at such small separation we could not resolve it from the quasar. The LBG in this field, detected here by MUSE, was not detected in the B band imagery for this field, likely due to the poor seeing and to the fact that that the B band is probing shorter wavelengths than the Lyman break of this galaxy.

\subsubsection{J0818+0720}

No high confidence or candidate LAEs are detected in association to DLA J0818+0720. In \cite{fumagalli2015} broad-band counterpart candidates to DLA J0818+0720 were detected in the WFC3/UVIS F390W imaging. These objects fall very close to the quasars ($0.7\pm0.1$ arcsec and $1.5\pm0.1$ arcsec), they appear to be compact, possibly stellar, in the WFC3 imaging. Even with the excellent seeing of the MUSE data these objects are too faint and too close to the quasar to have their spectra extracted cleanly.

\subsubsection{J0818+2631}

Two LAE candidates are detected in proximity to DLA J0818+2631, but there are no high confidence detections in this field. Id91 is a promising candidate for a direct association with the DLA, as it has a velocity offset from the DLA of just +290.5 km s$^{-1}$ and an impact parameter of 7.0 arcsec (52.5 kpc). The other candidate LAE in this field, id87, has a similar position in velocity space (+231.5 km s$^{-1}$) but has a larger impact parameter (170.1 kpc). Due to only 2 of 6 exposures for this field being taken the data quality is poor, and some objects are rejected because they fell in regions where one of the two exposures is masked, preventing us from confirming the robustness of these detections. 

The candidate host galaxy identified \cite{fumagalli2015} for this DLA is very close to the quasar line of sight (2.05 arcsec), but with MUSE this galaxy was confirmed to be a low redshift interloper at $z=0.554$, highlighting once more the need for spectroscopic searches for DLA host galaxies.

\section{Spectra of LBGs and candidate LAEs}\label{ap:LBGs}

\begin{figure*}
	\centering
	\includegraphics[width=1.0\textwidth]{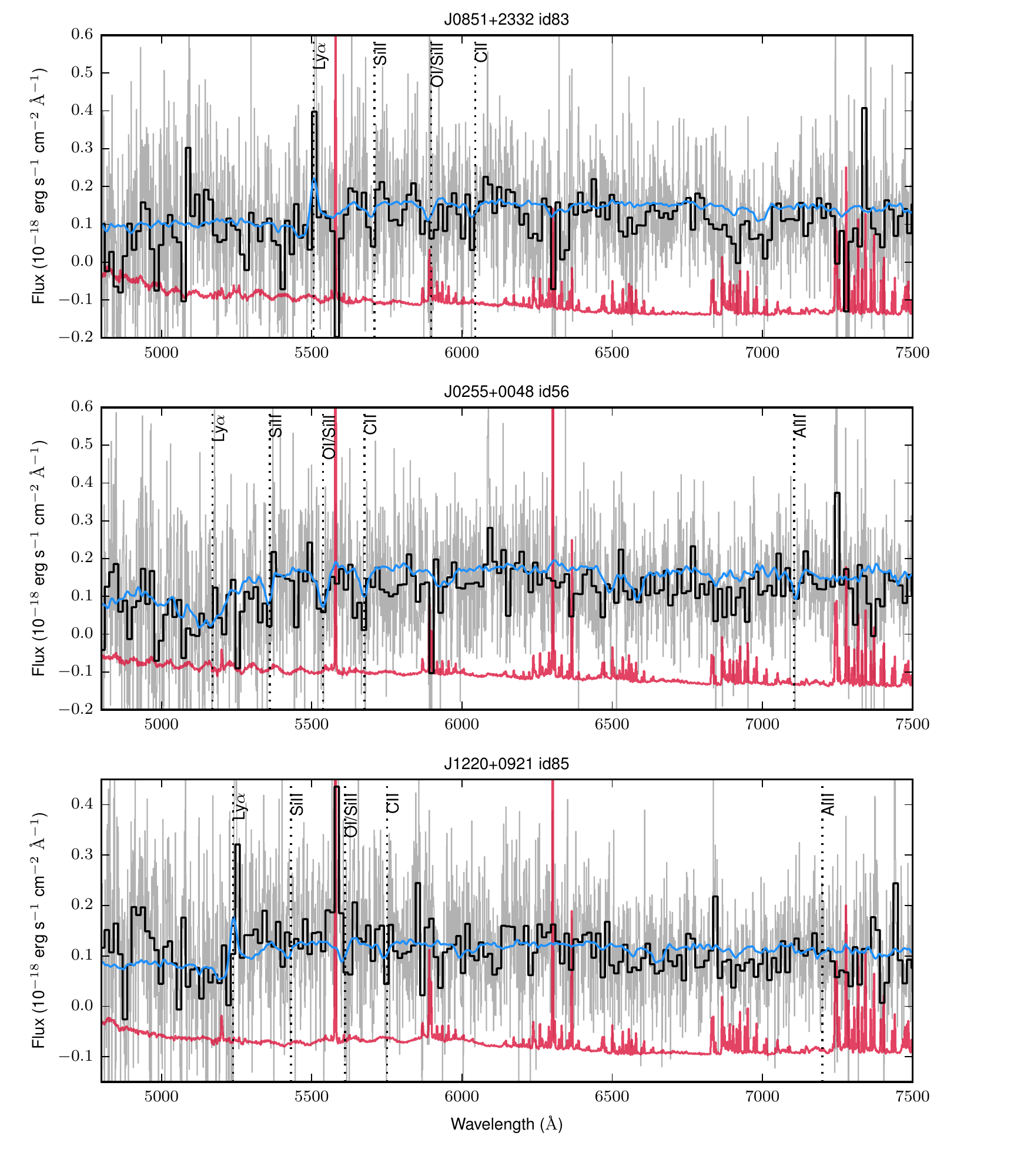}
	\caption{MUSE spectra of the detected continuum counterparts to LAEs in DLA fields. In each panel the 1D MUSE spectrum integrated over the object is shown (grey) over-plotted with a re-binned version of the same spectrum to suppress the noise (black). The 1 $\sigma$ flux error in the unsmoothed spectrum is shown with shifted down for clarity, with zero at the bottom of each plot (red). With each object a template LBG spectrum is plotted (blue) redshifted to match the sources, the templates shown were selected to roughly match the presence of \lya\ emission or absorption. The wavelengths of lines commonly observed in LBG spectra are also labelled (vertical dotted lines).}
	\label{fig:LBG_spec}
	\vspace{-1em}
\end{figure*}

\begin{figure*}
	\centering
	\includegraphics[width=1.0\textwidth]{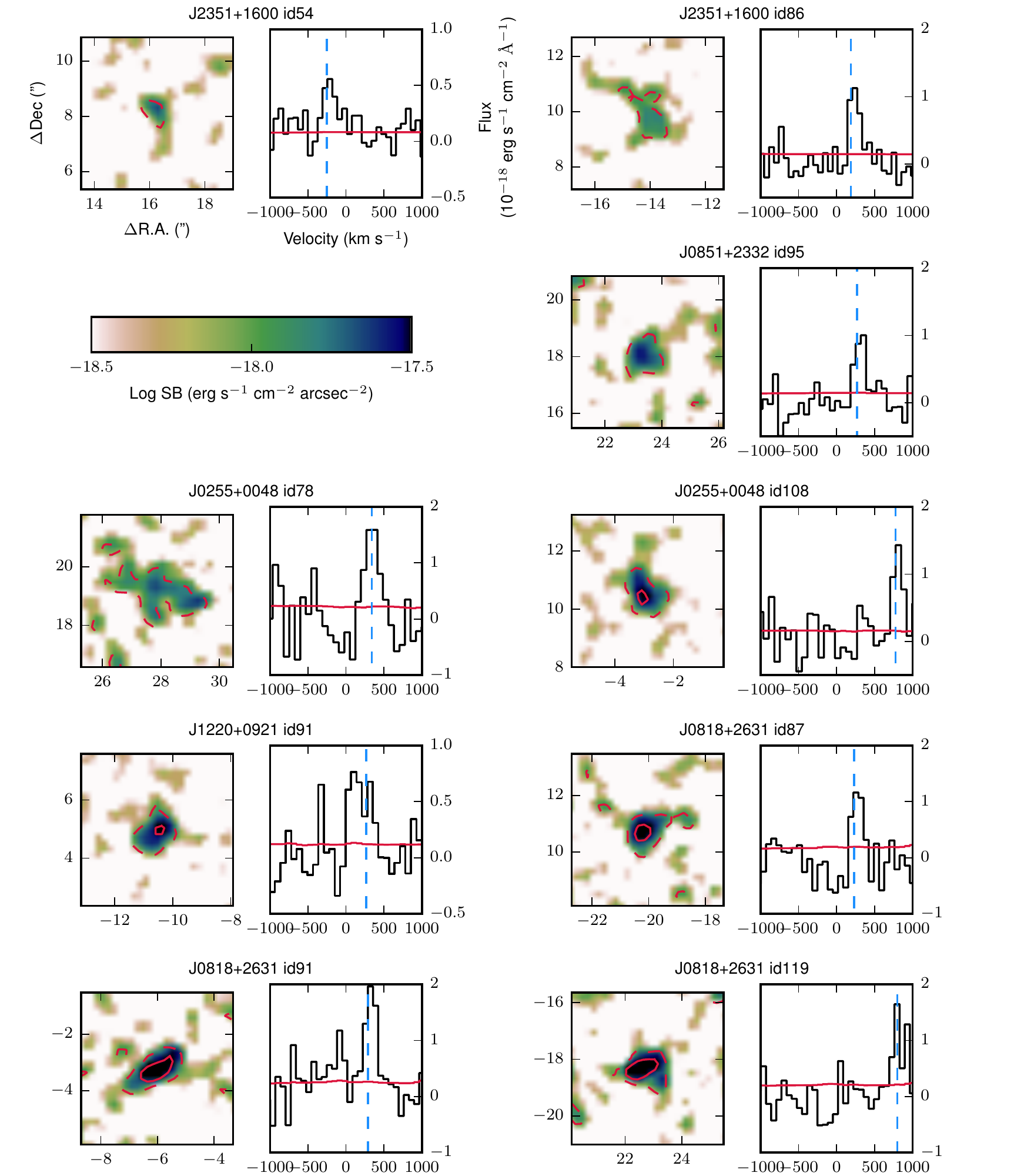}
	\caption{Cutouts of the lower SNR LAE candidates. Each object has an optimally extracted \lya\ surface brightness map (left) and an 1D spectrum of the detected emission line. The images have \lya\ SB contours (red), drawn at at 10$^{-18}$ (dashed) and 10$^{-17.5}$ (solid) erg s$^{-1}$ cm$^{-2}$ arcsec$^{-2}$. The \lya\ SB map has been smoothed with a Gaussian kernel ($\sigma=1$ pixel) and the axes are in physical kpc from the DLA. For the spectra the flux (black), 1$\sigma$ flux error (red) and line central velocity (vertical dashed blue line) and shown plotted as a velocity offset from the DLA redshift.}
	\label{fig:other_cands}
\end{figure*}

In this Section, we report the spectra of galaxies with detected continuum counterparts at the depth of our MUSE observations (Fig.~\ref{fig:LBG_spec}), as well as spectra and reconstructed line images of the candidate LAEs (Fig.~\ref{fig:other_cands}).

For the LAEs in the fields J0255+0048 and J1220+0921, the masks used to extract the LBGs spectra are taken from the {\sc SExtractor} segmentation masks as described above. In the case of the structure in field J0255+0048, the continuum counterpart is only $\simeq2.5$ arcsec from the quasar and hence, is not deblended  by SExtractor. For this object, the mask to extract the spectrum is generated manually, using a circular aperture with a diameter of 1.2 arcsec. Due to its proximity to the quasar, the spectrum is contaminated, and requires careful sky subtraction. An annual sky aperture around the quasar, but excluding the continuum counterpart is used to extract a sky spectrum, which is then used to subtract the quasar contamination from the spectrum.

\bsp	
\label{lastpage}

\end{document}